\documentstyle[editedvolume,psfig]{lipari}    

\def\etal  {{et~al.}\ }

\newcommand{\ga}{\stackrel{_{>}}{_{\sim}}}

\def\amin{\ifmmode ^{\prime}\else$^{\prime}$\fi}
\def\asec{\ifmmode ^{\prime\prime}\else$^{\prime\prime}$\fi}
\def\lta{\mathrel{\hbox{\rlap{\lower.55ex \hbox {$\sim$}}
        \kern-.3em \raise.4ex \hbox{$<$}}}}
\def\gta{\mathrel{\hbox{\rlap{\lower.55ex \hbox {$\sim$}}
        \kern-.3em \raise.4ex \hbox{$>$}}}}

\begin{opening} \title{NEUTRON STARS AND BLACK HOLES IN X-RAY BINARIES} 
\subtitle{}  

\author{Jan van Paradijs}  
\institute{Astronomical Institute ``Anton Pannekoek'', UvA, and 
CHEAF, Amsterdam, The Netherlands, 
\& 
Department of Physics, UAH, Huntsville AL 35899, USA}   

\end{opening}  

\begin{document}   

\begin{abstract} Galactic accretion driven stellar X-ray sources can be
divided into groups in different ways. An important division, which covers
almost all known X-ray binaries, can be made according to the mass of the
donor star: high-mass X-ray binaries and low-mass X-ray binaries.
Another distinction (partially overlapping with the previous one) can be
made on the basis of the nature of the accreting object: a strongly
magnetized neutron star, a neutron star with a weak magnetic field, or a
black hole. In this review I describe the properties of these different 
types of X-ray binaries, and discuss the mass determinations which are the
basis for distinguishing accreting neutron stars from black holes.
\end{abstract}  

\section{Some Historical Background}   

The connection between cosmic X-ray sources and compact
stars is an old one: soon after the discovery of the first such
source, Sco X-1 (Giacconi et al. 1963), it was proposed that these
objects are young hot neutron stars, formed in recent supernovae, which
cool through thermal radiation from their surfaces (Chiu 1964; Chiu \&
Salpeter 1964; Finzi 1964). However, the finite extent of the X-ray
source associated with the Crab Nebula (Bowyer et al. 1964), and the
non-Planckian shape of the X-ray spectra of this source (Clark 1965)
and of Sco X-1 (Giacconi et al. 1965) showed that this was not, in
general, a good model for X-ray sources.    

Accretion onto a compact star had meanwhile been suggested as a
possible source of energy for quasars and X-ray sources (Salpeter
1964; Zel'dovich 1964; Zel'dovich \& Guseinov 1964), and together with
the peculiarities of the optical spectra of the counterparts of Sco
X-1 (Sandage et al. 1966) and Cyg X-2 (Giacconi et al. 1967) this led to
the idea that these sources are mass-exchanging binaries with a
compact component (Shklovsky 1967; see also Burbidge 1972; Ginzburg
1990). The spectrum of Sco X-1 was similar to those of old novae and U
Gem type stars, which were by then known to be binary stars, in
particular through the work of Crawford and Kraft  in the 1950s and
'60s (Crawford \& Kraft 1956; Kraft 1962, 1964).  The optical spectrum
of Cyg X-2 was found to be composite, showing the signatures of both a
late-type star and a component of much higher excitation; Cyg X-2 also
showed significant radial-velocity variations.  However, the
single-most important characteristic of a binary star, i.e., an
orbital periodicity, was not found in either system  until many years
later, in spite of substantial observational effort (see Hiltner \&
Mook 1970, and Kraft 1973, for discussions of early optical
observations of X-ray sources, and references).   

\begin{figure}
\centerline{
\psfig{figure=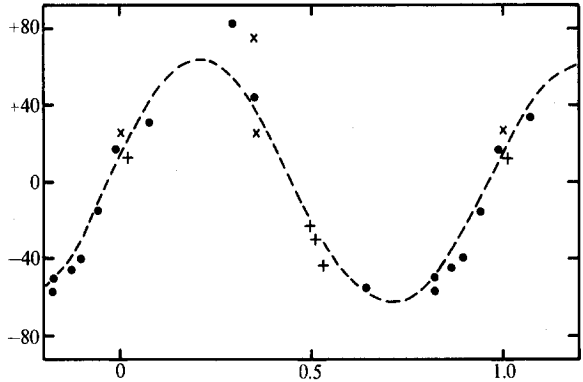}}
\caption{Radial-velocity curve of HD~226868, the optical counterpart of
Cyg X-1, providing evidence that this source is a binary star with an
accreting black hole (from Webster \& Murdin 1972).}
\end{figure}

The discovery of the first X-ray binary, Cyg X-1, by Webster \& Murdin
(1972) and Bolton (1972) followed the determination of an accurate
($\sim 1^{\prime}$) error box by Rappaport  et al. (1971), which
contained a radio source (Braes \& Miley 1971;  Hjellming \& Wade
1971) coincident with the 8th magnitude known supergiant HD 226868. 
Optical spectroscopy of this  star revealed a 5.6 day periodic
radial-velocity variation with an  amplitude $K_{\rm opt} = 64$ km/s,
and a corresponding mass function (see Sect. 4.3.1) $f_{\rm opt}(M) =
0.25$~M$_{\odot}$. Under the assumption that the  supergiant has a
``normal'' mass of $\gta 15$~M$_{\odot}$ these results  led to the
conclusion that the mass of the compact star in Cyg X-1 is higher
than 3~M$_{\odot}$, which exceeds the maximum possible mass  of a
neutron star. Thus, the first X-ray source for which the binary nature
was established, contained an accreting object that is likely to be a black
hole. (See Section 3.1 for a discussion of the term ``black
hole'' in the context of X-ray binaries). 

The identification of the radio source with Cyg X-1 was confirmed when
its brightness showed a large increase
correlated with a major hardening  of the 2-10 keV spectrum of Cyg X-1
(Tananbaum et al. 1972).   We now know that this spectral hardening is
caused by the disappearance of  an ``ultra-soft'' spectral component in
the X-ray spectrum, signalling  a transition from a ``high'' (or ``soft'')
state to a ``low'' (or ``hard'') state  (see Sect. 3.3).

The idea that all bright galactic X-ray sources are mass-exchanging
binary stars with a compact accretor became widely 
accepted with the observation of regular 
eclipses of the pulsating X-ray source Cen X-3 (Giacconi et al. 
1971; Schreier et al. 1972).  
The variable delays of the pulse arrival times, in 
phase with the periodic (2.1 days) eclipses of the X-ray source,
showed persuasively that in Cen X-3 the X rays are generated by
accretion onto a strongly magnetized neutron star, rotating at a
4.8 s pulse period, in orbit around a massive ($\ga$ 10~M$_{\odot}$)
companion star.  This was later confirmed by the optical identification of
this source with an O-type giant star (Krzeminski 1974).  The discovery
of the binary nature of Cen X-3 was soon followed by more
observations of eclipsing X-ray sources, some of them pulsating, and
by the
identification of these X-ray sources with early-type stars (see, e.g.,
Liller 1973; Penny et al. 1973; Vidal 1973). In addition, a general
framework for the origin and evolution of a massive X-ray binary, as a
rather normal episode in the life of a massive close binary star with
successive stages of mass transfer between the two components, was
readily accepted (Van den Heuvel \& Heise 1972).  Thus, within a few years
the existence of a galactic population of high-mass X-ray binaries (HMXB),
with accreting neutron stars (or occasionally a black hole) was well
established.  

\begin{figure} 
\centerline{
\psfig{figure=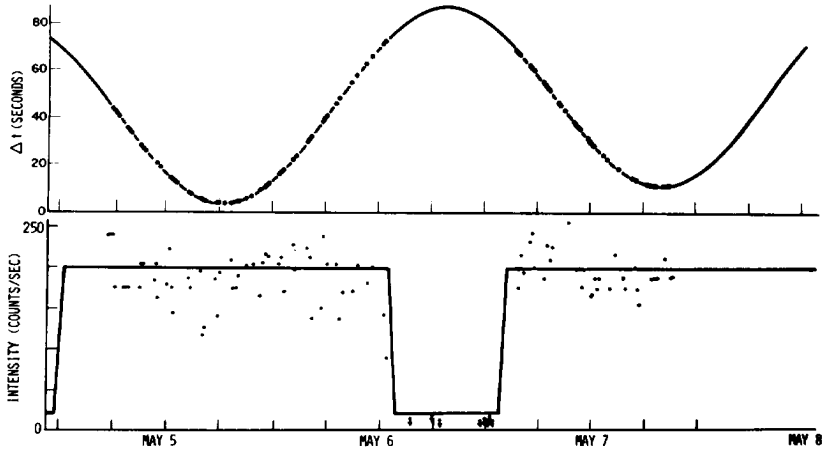}}
\caption{The top panel shows the Doppler delay of the arrival times of the
4.8 s X-ray pulsations of Cen X-3 (relative to the times expected for a
constant pulse period), as observed with Uhuru in May 1971. The bottom
panel shows the X-ray intensity variations during the same time. The
maximum delay coincides with the center of the X-ray eclipse (adapted
from Schreier et al. 1972).} 
\end{figure}  

Already in the 1960's (see, e.g., Dolan 1970) it had become clear that
there is a clustering of bright X-ray sources within $\sim 30^{\circ}$
of the direction of the galactic center. This concentration was not
accompanied by a strong background of unresolved sources, which showed
that these sources are located in the central regions of the Galaxy
(Ryter 1970; Setti \& Woltjer 1970). It was, therefore, suspected that
apart from the above-described HMXB there is a class of low-mass X-ray
binaries (LMXB) with donor star masses of the order of a Solar mass or
less (see, e.g., Salpeter 1973), but proof for this idea was hard to
obtain.  Apart from the difficulty of finding orbital periods, the
apparent heterogeneity of the properties of LMXB may have played a
role.  Compared to the HMXB the first handful of systems now
classified as LMXB (Her X-1, Sco X-1, Cir X-1) show rather more
diversity than similarity in their properties.  As a result, 
only at the end of the 1970s did it become clear that there are
such objects as low-mass X-ray binaries, which form a group
with ``family traits'',  distinct from the HMXB with respect to their
sky distributions, X-ray spectral characteristics, optical properties,
and types of X-ray variability (see, e.g., Lewin \& Clark 1980). The LMXB
comprise the globular-cluster X-ray sources, X-ray bursters, soft
X-ray transients, and the bright galactic-bulge X-ray sources. The
properties of HMXB and LMXB are discussed in Section 2. 

Roughly speaking, the reason for the bi-modal distribution of the masses
of donor stars in X-ray binaries is that for stars less massive than 
$\sim10$~M$_{\odot}$ the stellar wind is too weak to power a strong X-ray
source; on the other hand, Roche lobe overflow is unstable for 
stars more massive 
than a neutron star, and proceeds on a very short time scale, as a
consequence of which the accreting neutron is completely engulfed and X
rays cannot escape.   

The differences between the X-ray properties of  LMXB and HMXB (with
accreting neutron stars) may be linked to a difference in the strength
of the magnetic fields of the neutron stars they harbour. The natural
assumption that the difference in donor star masses corresponds to a
difference in the ages of LMXB and HMXB has led to the idea that the
magnetic fields of neutron stars decay with time. This topic is
briefly discussed in Section 2.6.

For a relatively small, but increasing, number of X-ray binaries (both
HMXB and LMXB) there is evidence that the accreting object is a black
hole. This evidence is based on mass determinations of the compact
star from the orbital parameters of the X-ray binary. 
In Section 3 I discuss the X-ray properties of 
accreting black holes, and attempts that have been made to recognize
X-ray characteristics that might allow one to ``easily'' identify black
holes in X-ray binaries. An overview of mass
determinations of neutron stars and black holes is given in
Section 4. 

Within the limits of this review I cannot strive for completeness. 
For detailed reviews on a variety of topics related to X-ray binaries
I refer the interested reader to the recent book ``X-ray Binaries''
(Lewin, Van Paradijs \& Van den Heuvel 1995). Further background
information can be found in individual chapters in the books by
Shapiro \& Teukolsky (1983),  Frank, King \& Raine (1992), {\"O}gelman
\& Van den Heuvel (1989), Ventura \& Pines (1991), Van den Heuvel \&
Rappaport (1992), and Alpar, Kiziloglu \& Van Paradijs (1995).
References on individual sources can be found in Bradt \& McClintock
(1983) and Van Paradijs (1995). An extensive summary of X-ray
satellite missions has been given by Bradt, Ohashi \& Pounds (1992). 

\section{High-Mass X-ray Binaries and Low-Mass X-ray Binaries}

\subsection{ Optical Counterparts }  

\subsubsection{High-mass X-ray binaries }  

The optical counterparts of HMXB have normal early-type spectra, in the
sense that these can be MK-classified (cf. Kaler 1989) 
without particular difficulty, on the
basis of ratios of spectral line strengths (see Fig. 3). This is highly
informative, since it immediately provides us with at least an
approximate idea about the masses, radii, and ages of these stars. Some
disturbance of the spectrum, indicative of anisotropic gas flow near the
primary may show up as variable emission/absorption components,
particularly in H$\alpha$, H$\beta$, He II $\lambda$4686, and the 
C\,III-N\,III $\lambda$4630-50 complex. However, when the latter two lines are
strongly in emission (see e.g.,  Hensberge \etal 1973) this is likely due to
a very high temperature and luminosity 
of the primary (e.g., Of characteristics), and not to
the presence of the X-ray source. Except for the strong resonance
lines of abundant ions, the same is true for the UV spectra of HMXB. 
The reason that the X-ray source does
not seem to affect the spectral properties of the primary much, is that
the bolometric luminosity of the latter generally exceeds the X-ray
luminosity, often by a large margin (Van Paradijs \& McClintock 1995).   

\begin{figure} 
\centerline{
\psfig{figure=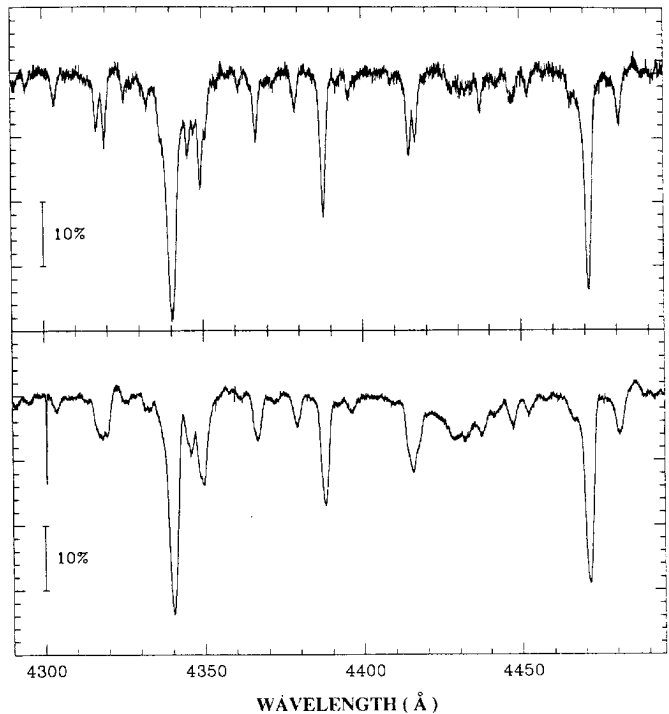}}
\caption{Comparison of the optical spectra of HD 77581, the optical
counterpart of the HMXB Vela X-1 (lower panel), and of the normal star 
$\kappa$ Ori (upper panel), which have spectral types B0.5 Ib and B0.5 Ia, 
respectively. Note the difference in line widths, caused by the larger 
rotational velocity of HD 77581 (courtesy  M. van Kerkwijk).} 
\end{figure}  

With respect to the spectral types of their optical counterparts the HMXB
can be divided into two subgroups, as follows. (i) The spectral type is
earlier than B2, and the luminosity class is I to III, i.e., the primary star
has evolved off the main sequence. The orbital periods are generally less
than $\sim$10 days. The donor stars fill, or almost fill, their Roche lobes,
as is apparent from the amplitudes of their optical light curves. (ii) The
primary is a B-emission (Be) star, characterized by emission lines (mainly
the Balmer series) which originate in circumstellar material. In the
Hertzsprung-Russell diagram they lie rather close to the main sequence.
The orbits of these Be/X-ray binaries are eccentric, and their periods tend
to be long. The primaries underfill their Roche lobes.   

As first suggested by Maraschi et al. (1975) the mass transfer in
these two groups is driven by different mechanisms. In the first group
mass is transferred via a strong stellar wind (in some sources a short-lived 
phase of Roche lobe overflow is observed). In the Be/X-ray binaries
the mass transfer is related to the anisotropic (often highly
variable) shedding of mass in the equatorial plane. This is observed
in all Be stars, and is  related to  their rapid rotation (Slettebak
1987; Briot 1986;  Bjorkman \& Cassinelli 1993). This inferred
difference in mass transfer mechanism is supported by the different
relations between orbital period and X-ray pulse period, first pointed
out by Corbet (1984), for these two groups of sources (see the
contribution by M. Finger to this Volume).

\begin{figure} 
\centerline{
\psfig{figure=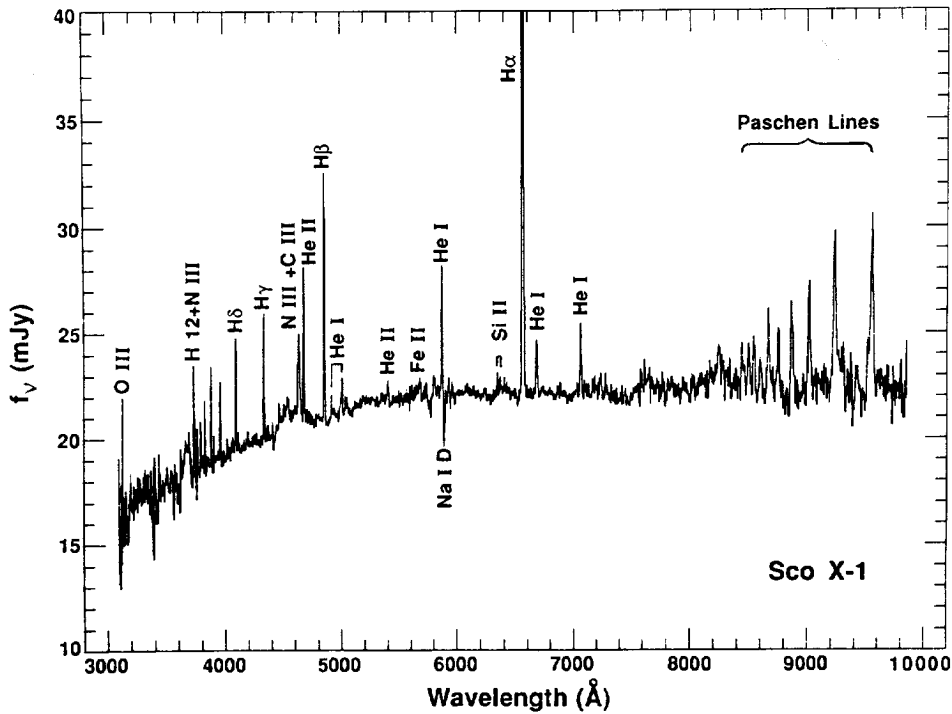}}
\caption{Optical spectrum of the low-mass X-ray binary Sco X-1
(from Schachter \etal 1989).} 
\end{figure}  

Most Be/X-ray binaries are highly variable, or transient. In some of
them recurrent outbursts have been observed, which reflect the varying
accretion rate onto the neutron star as it moves in its eccentric
orbit through regions of varying density around the Be star. In
addition, a more sudden turning on and off of the accretion can occur
when the wind density becomes too low for the neutron star
magnetosphere to be within the corotation radius, so that accretion
becomes centrifugally inhibited (Stella \etal 1986). However, in many
cases outbursts have been observed which are not related to the orbit
of the Be/X-ray binary. These outbursts are generally believed to be
due to a sudden enhancement of the mass loss of the Be star (for
reviews of various aspects of Be stars see Slettebak \& Snow 1987, and
Waters 1989); however, it has recently been suggested that these
outbursts may be caused by instabilities in an accretion disk around the
neutron star, analogous to those in dwarf novae and soft X-ray
transients (see M. Finger, this Volume.)

Many HMXB with evolved mass donors show optical brightness variations,
with two maxima and two minima per orbital cycle, which occur at
quadratures and conjunctions, respectively. These so-called ellipsoidal
variations are caused by the tidal and rotational distortion of the mass
donor which fills (or nearly fills) its Roche lobe, and the non-uniform
distribution of its surace brightness (see Sect. 4.3.2). 
In several cases the light curves are
also affected by X-ray heating of the mass donor, and by the presence of an
accretion disk (in LMXB these effects completely dominate the optical
light curves). Generally, in Be/X-ray binaries irregular variations, related
to their equatorial mass loss, dominate any orbital variability. 
For extensive discussions of the optical light curves of HMXB, see Tjemkes 
et al. (1986), Van Paradijs (1991) and Van Paradijs \& McClintock (1995).   

\subsubsection{ Low-mass X-ray binaries}  

The optical spectra of LMXB (e.g., Shahbaz et al. 1996a) 
show a few emission lines, particularly
H$\alpha$, H$\beta$, He\,II $\lambda$4686, and C\,III-N\,III 
$\lambda$4630-50, superposed on a rather flat (in frequency) continuum
(see Fig. 4). In very few cases the signature of a companion star can be
discerned. According to Motch \& Pakull (1989) the relative strength of
the C\,III-N\,III emission complex relative to the $\lambda$4686 emission
provides a good measure of the heavy-element abundances in the accreted
matter. Spectra of cataclysmic variables (CVs), in which the accretor is
not a neutron star (or black hole) but a white dwarf, bear a general
resemblance to those of LMXB, showing emission lines superposed on a
continuum (see Warner 1995, for a comprehensive review of CVs). 
However, the equivalent widths of these lines in LMXB spectra,
in particular that of H$\beta$,  tend to be much smaller than those in CV
spectra (Van Paradijs \& Verbunt 1984; Shahbaz et al. 1996a).    

In LMXB the donor star transfers mass by Roche lobe overflow. This mass
arrives at the compact star via a relatively flat rotating 
configuration, the accretion disk, in which it slowly spirals inward. 
LMXB spectra originate from this accretion disk, which radiates mainly
through reprocessing of incident X-rays into optical/UV photons. This
reprocessing dominates the internal energy generation of the disk due
to  conversion of gravitational potential energy into heat,  generally
by a large factor (see Van Paradijs \& McClintock 1995). For internal
energy generation alone, as occurs in accretion disks in cataclysmic
variables (CVs), the (local) effective temperature, $T$,  in the disk
varies with radial distance, $r$,  from the central compact star 
approximately as
$T(r) \propto r^{-3/4}$. In LMXB disks, where reprocessing of X rays
dominates the energy budget, one would expect $T(r) \propto r^{-1/2}$. 
Thus, the
farther out in the disk, the more reprocessing dominates internal
energy generation. According to the simplified, but self-consistent 
calculations of Vrtilek et al. (1990), in which also the effect of X-ray 
irradiation on the height of the disk is taken into account, $T(r) \propto 
r^{-3/7}$. 

Many LMXB show a regular  orbital variation of their optical brightness,
with one maximum and minimum per orbital cycle. Minimum light occurs
at the superior conjunction of the X-ray source. These variations are
caused by the changing visibility of the accretion disk and the X-ray
heated side of the secondary, insofar as the latter is not in the X-ray
shadow of the disk (see Van Paradijs 1991, and Van Paradijs \& McClintock 
1995, for reviews of the optical light curves of LMXB).

For transient LMXB (``soft X-ray transients'', see Sect. 3.4) in quiescence 
X-ray heating of the accretion disk and the companion star is not very 
important, and the optical emission of the system is dominated by the Roche 
lobe filling secondary star. These systems then show ellipsoidal light 
curves (see also Sect. 4.3.2).

The (reddening-corrected) colour indices $B-V$ and $U-B$ of LMXB have
average values of $-0.09 \pm 0.14$ and $-0.97 \pm 0.17$, respectively
(errors are $1 \sigma$ standard deviations), close to those expected
for a flat continuum ($F_{\nu}$ = constant).  The distribution of the
ratio of X-ray to optical fluxes is rather sharply peaked.  Expressed
in terms of an ``optical/X-ray colour index'' $B_{0}$ + 2.5 log
$F_{\rm X}$($\mu$Jy), the peak occurs near 21.5, corresponding to a
ratio of fluxes emitted in X rays (2-11 keV) and in optical light
(3000-7000 \AA) of $\sim 500$ (Van Paradijs \& McClintock 1995).   

The optical luminosities of LMXBs are, in general, much higher than
those of CVs (see, e.g., Warner 1987, 1995); this is because the
gravitational potential well of a neutron star  is much deeper than
that of a white dwarf, and X-ray heating of the accretion disk is not
important in CVs.   

Absolute visual magnitudes $M_{\rm V}$ have been estimated for LMXB with 
distance estimates: (i) LMXB in stellar systems at a known
distance; (ii) X-ray burst sources showing bursts with photospheric radius
expansion; during this phase of expansion the burst luminosity is very
close to the Eddington limit, which is a reasonably good standard candle
(see Lewin, Van Paradijs \& Taam 1993; 1995). Thus, the X-ray luminosity (in
units of the Eddington limit) is given by ratio $\gamma$ of the persistent
flux to the peak flux of radius expansion bursts; (iii) Z sources (see sect.
2.4.4); when these are in the ``normal-branch'' state their X-ray luminosity is
very close to the Eddington limit (see, e.g., Van der Klis 1995); (iv) Soft
X-ray transients, i.e., LMXB whose mass accretion rate is high only during
rather brief time intervals, and very low during most of the time. Their
distance can be determined from the spectral properties of the companion
star, which becomes detectable during ``quiescence'' when reprocessing of
X rays in the accretion disk is not important. The absolute magnitudes of
active LMXB range between $-5$ and $+5$ (Van Paradijs \& McClintock
1994).
This large range is the consequence of the large range in X-ray luminosity,
$L_{\rm X}$, of the central source, and in the size of the accretion disk. For a
simple model of reprocessing of X rays the optical luminosity, $L_{\rm
V}$,  of the disk is expected to scale with $L_{\rm X}$ and orbital period P
as $L_{\rm V} \propto  L_{\rm X}^{1/2}P^{2/3}$, in agreement with the
values of $M_{\rm V}$ for LMXB with known orbital periods (See Fig.
5). This is confirmed by numerical calculations of X-ray heated accretion 
disks (De Jong et al. 1996).

\begin{figure}
\centerline{
\psfig{figure=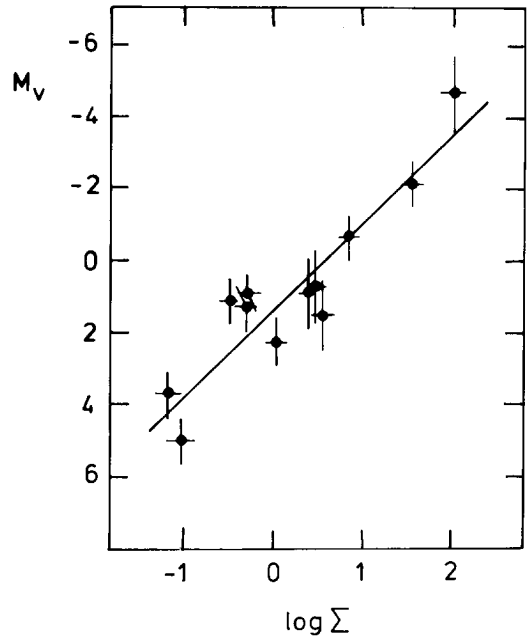}}
\caption{Relation between the absolute magnitude $M_{\rm V}$ and $\Sigma =
L^{1/2}_{\rm X}P^{2/3}$ (from Van Paradijs \& McClintock 1994)}
\end{figure}

\subsection{Orbital Periods} 

\subsubsection{High-mass X-ray binaries} 

Orbital periods have been determined for 28 HMXB using
various methods, e.g., X-ray eclipse timing, Doppler delay curves of
X-ray pulse arrival times, and optical brightness and radial-velocity
variations; they range between 4.8 hours and 188 days (Van Paradijs
1995; see also Table 1).  
The orbital periods are correlated with the
type of mass transfer that powers the X-ray source, and with the
evolutionary state of the binary system. The shortest orbital period
by far (4.8 hours) is that of Cyg X-3, whose mass donor is a helium
star (Van Kerkwijk \etal 1992, 1996). This source represents a late
stage in the evolution of a ``normal'' X-ray binary, which has passed a
common-envelope phase in which the primary has lost its hydrogen-rich
envelope. Systems with (incipient) Roche lobe overflow (e.g., LMC X-4,
LMC X-3, Cen X-3), tend to have orbital periods of a few days. (Once
Roche lobe overflow has fully developed the mass tranfer rate will
become very large, and the compact star is expected to be completely
shrouded and not to be detectable as an X-ray source.) Somewhat longer
periods are found for the HMXB with evolved (supergiant) companions
which transfer mass via a strong stellar wind (e.g., 1700$-$377,
1538$-$522, Vela X-1). The HMXB with the longest orbital periods (up
to hundreds of days) have Be star primaries (see Section 2.1.1). These
have very eccentric orbits, which can only be explained if neutron
stars get a kick velocity at their formation (Van den Heuvel 1994; Verbunt 
\& Van den Heuvel 1995).

\begin{table}
\caption{ X-ray binary Orbital Periods (determined since 1993)}
\begin{center}
\begin{tabular}{lcc}
\hline \\
LMXB & $P_{\rm orb}$ (hours)  & References \\ \hline
\\ 
GRO~J0422+32   &  5.1  & [1] \\
GRS~1009--45   &  6.9  & [2] \\
GRO~J1655--40 & 62.9  & [3,4] \\
4U~1702--363   & 22.3 or 358 & [5,6,7] \\
H~1705--25     & 12.5  & [8] \\
GRS~1719--249  & $\sim 14.7$ & [9] \\
A~1742--289     &  8.4  & [10] \\
GRO~J1744--28  & 283   & [11] \\ 
4U~1850--087   & 0.34  & [12] \\ \hline
\\ 
HMXB & $P_{\rm orb}$ (days)  & References \\ \hline
\\
RX~J0648.1--4419 &  1.54 & [13] \\
GRS~0834--430   &  10.6 & [14] \\
GRO~J1008--57   & $\sim 248$ &[15] \\
2S~1417--624    & 42.1  & [16] \\
GRO~J1750--27   & 29.8  & [17] \\ \hline

\end{tabular}
\end{center}

\footnotesize
[1] Chevalier \& Ilovaisky (1996);
[2] Shahbaz et al. (1996b);
[3] Orosz \& Bailyn (1997);
[4] Van der Hooft et al. (1997);
[5] Wachter \& Margon (1996);
[6] Southwell et al. (1996); 
[7] Barziv et al. (1997); 
[8] Remillard et al. (1996); 
[9] Masetti et al. (1996);
[10] Maeda et al. (1996); 
[11] Finger et al. (1996a);
[12] Homer et al. (1996);
[13] Israel et al. (1995);
[14] C.A. Wilson et al. (1997);
[15] R.B. Wilson et al. (1996);
[16] Finger et al. (1996b);
[17] Scott et al. 1997.
\normalsize
\end{table}

\subsubsection{Low-mass X-ray binaries} 

Orbital periods are known for some three dozen LMXB, mainly from regular X-ray
eclipses and X-ray ``dips'' (White \etal 1995), and optical brightnes
variations (Van Paradijs \& McClintock 1995). In several cases
periodic radial-velocity variations have been found for emission lines
(originating from an accretion disk), or for absorption lines
(secondary star of soft X-ray transients in quiescence). The known
orbital periods of  LMXB range between 685 seconds and 16.6 days (Van
Paradijs 1995; see also Table 1). This large range indicates that LMXB
have a variety of mass donors  (white dwarfs, main-sequence stars,
giant stars). The orbital-period distributions of CV and LMXB (Fig.
6) are different. Compared to the CVs a much larger fraction of LMXB
have periods above about half a day. A possible explanation for this
difference is that for such long periods the companion star masses are
expected to substantially exceed the mass of a typical white dwarf, 
but not that of a neutron star. This will make the mass transfer
at long periods unstable for CVs, but not for LMXB.  The LMXB period
distribution does not show the well-known period gap in the
distribution for the CVs (see, e.g., Verbunt 1984, Spruit \& Ritter 1984, 
for studies of the
period gap); in fact, there are no LMXB in the period range between 
80 minutes and 2 hours (i.e., below the period gap), which is well
populated by the CV. This may be the result of  the high probability
that LMXB form at periods of about half a day or longer, wheres a
large fraction of CVs form at periods below 2 hours (King \& Kolb 
1997). Perhaps evaporation of the LMXB secondaries plays a role after
the LMXB has reached the upper edge of the period gap; mass transfer
then stops, and the rapidly rotating neutron star (spun up by
accretion torques) then becomes active as a millisecond radio pulsar
(Ruderman \etal 1989; Van den Heuvel \& Van Paradijs 1988); however,
whether or not complete evaporation of the secondary star occurs, is a
matter of debate (see, e.g., Emmering \& London 1990).   

\begin{figure} 
\centerline{
\psfig{figure=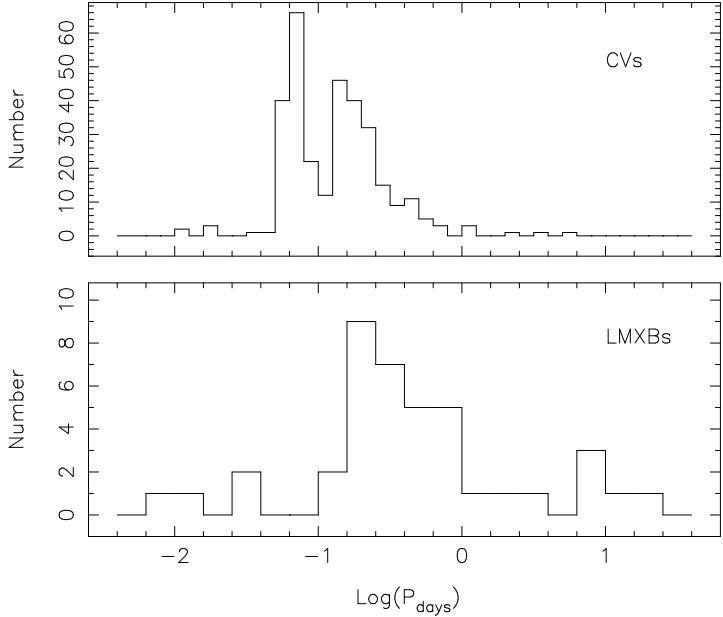,width=12cm,angle=-90}}
\caption{Distributions of orbital periods for
low-mass X-ray binaries and cataclysmic variables. Data have been taken
from Van Paradijs (1995), Ritter \& Kolb (1997), and the references
given in Table 1.} 
\end{figure}  

\subsection{Galactic Distributions of HMXB and LMXB} 

The sky distributions of the HMXB and LMXB are shown in Fig. 7. The
galactic HMXB are distributed along the galactic plane, without an obvious 
concentration to the galactic center. They have an
average latitude $<b^{\rm II}> = -0.8 \pm 6.9^{\circ}$; if we leave
out X Per and a few other nearby high-latitude Be/X-ray systems
identified by Tuohy \etal (1988) we find $<b^{\rm II}> = 0.4 \pm
1.9^{\circ}$.  This fits the idea that HMXB are young population I objects.
The HMXB population is  dominated by the Be/X-ray binaries. Based on
evolutionary considerations Meurs \& Van den Heuvel (1989) estimated 
that there are (2-6) $10^{3}$ of them in the Galaxy; the majority of
them are inactive. This is consistent with the empirical estimate of 
Bildsten et al. (1997), who find the total number of Be/X-ray
transients in the Galaxy to be in the range $10^2$ to $\sim 10^3$. The
total number of galactic HMXB with an evolved component is $\sim$40
(Meurs \& Van den Heuvel 1989; Van Paradijs \& McClintock 1995). 

A statistical study of the kinematic properties of the optically
identified HMXB (Van Oijen 1989) indicates that these objects are
runaway stars. In particular, they appear not to be members of OB
associations. This was recently confirmed beautifully with the
detection of a bow shock around Vela X-1, the result of the latter's
motion through the interstellar medium with a total velocity of $\sim
90$ km/s (Kaper et al. 1997).

The galactic LMXB (excluding the globular-cluster sources) have a
wider latitude distribution ($<b^{\rm II}> = 0.4 \pm 9.1^{\circ}$), 
and are also more concentrated to the direction of the galactic
center. The scale height of the (assumed exponential) distribution of
distances from the galactic plane, for LMXB with neutron stars at 
known distances (see Sect. 2.1.2) is ~900 pc (Van Paradijs \& White
1995). The $z$ distribution of LMXB with a black  hole is
substantially narrower (White \& Van Paradijs 1996). The wide $z$
dispersion of LMXB with neutron stars requires that the neutron
stars in these systems formed in an asymmetric supernova explosion
which gave them an extra kick velocity (Brandt \& Podsiadlowski 1995;
Van Paradijs \& White 1995; Ramachandran \& Bhattacharya 1997; see
also Ramachandran's contribution to this Volume); the kick velocity
distribution is consistent with that of single radio pulsars (Lyne \&
Lorimer 1994; Hansen 1996; Hansen \& Phinney 1997; Hartman 1997).

There are $\sim 10^2$ persistent luminous LMXB ($L_{\rm X} > 10^{36}$
erg/s) in the Galaxy (Van Paradijs 1995). During the last decade it has
become clear that in a large fraction of the transient LMXB the
compact star is a black hole (see Section 3.4); the number of such 
transient black-hole binaries in the Galaxy is estimated to be between
a few hundred and a few thousand (Tanaka \& Lewin 1995; White \& Van
Paradijs 1996).

The kinematic properties of LMXB have been studied by Cowley et al. 
(1988) and Johnston (1992). Based on the high velocity dispersion and
low galactic rotation velocity Cowley et al. (1988) concluded that
LMXB are among the oldest objects in the Galaxy. However, since LMXB
get a kick velocity at the formation of the neutron star, such a
direct interpretation of the kinematic properties of LMXB in terms of
their ages cannot be made.

\begin{figure} 
\centerline{
\psfig{figure=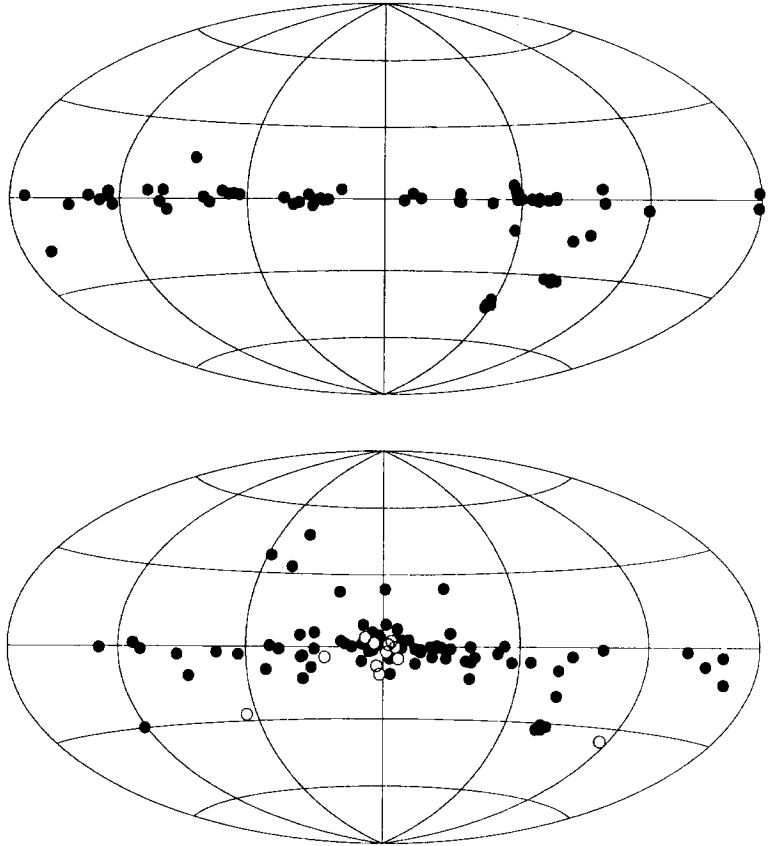}}
\caption{Sky maps (in galactic coordinates) of the high-mass X-ray
binaries (top panel) and low-mass X-ray binaries (bottom panel); the
latter also include the globular-cluster sources (indicated by open 
circles). The 27 LMXB within 2$^{\circ}$ of the Galactic center have not 
been included to avoid congestion of the map. These maps are based on the 
catalogue of Van Paradijs (1995).} 
\end{figure}  

\subsection{X-ray Variability} 

\subsubsection{X-ray pulsations}   

Almost all HMXB show X-ray pulsations, which indicates that the
accreting compact stars in these systems are strongly magnetized neutron
stars (for reviews of various aspects of X-ray pulsars see, e.g., Joss \&
Rappaport 1984; Nagase 1989; White et al. 1995; Bildsten et al. 1997). 
Strong magnetic fields
(a few $10^{12}$ G) have also been inferred from the presence of cyclotron
lines in the hard X-ray spectra of some X-ray pulsars (see Mihara 
et al. 1991, for references).

Observed pulse periods range over a factor $\sim10^{4}$, between 69
msec and 24 minutes. A statistical analysis of pulse profiles of
different  sources, and  observations of the transient source 
EXO~2030+475 indicate that pulse profiles are correlated with the
source luminosity, with the emission pattern changing from a fan beam
to a pencil beam pattern as the luminosity decreases (White et al.
1984; Parmar et al. 1989). 

The location of X-ray pulsars in a spin period versus orbital period diagram 
(Corbet et al. 1984) is correlated with the properties of the companion 
star; evolved massive companions can be distinguished in 
this diagram from those that are not evolved, as can low-mass from 
high-mass companions, and high-mass 
companions which fill their Roche lobe from those that don't (see also M. 
Finger's contribution to this Volume).

Pulse arrival time measurements for pulsating HMXB, in combination with
radial-velocity observations of their massive companions, have provided
information on the masses of accreting neutron stars. This is discussed
in Section 4. 

X-ray pulse periods show long-term changes which are caused by a
combination of external accretion torques, and internal torques due to
the coupling of the crust and the core of the neutron star (see, e.g.,
Henrichs 1983). Based on observations made before 1990 it appeared
that sources in which the accretion takes place via an accretion disk
show a fairly regular long-term spin up of the pulsar, with typical
time scales of roughly $10^{4}$ years, and that sources with
stellar-wind accretion show much more irregular spin period
fluctuations (see Lamb 1988, 1989; Nagase 1989, for detailed reviews,
and references). However, the continuous monitoring of X-ray pulsars
with BATSE since 1991 has shown that this result reflects the sparse
sampling of pulse period measurements: the disk accretors show
frequent torque reversals, with a bi-modal distribution of spin up/down
rates. This is discussed in detail in the contribution of M. Finger to
this Volume.

There is an interesting group of X-ray pulsars, distinguished by pulse
periods in a narrow range near 6 s, relatively soft X-ray spectra, low
luminosities, and a tendency to spin down regularly. Optical
counterparts have not been found so far. These objects, which may be
the relatively recent result of a common-envelope evolution of massive
binary stars, are discussed in the contribution to this Volume by
Stella, Israel \& Mereghetti.

\subsubsection{X-ray bursts}   

Many LMXB emit X-ray bursts, during which the X-ray flux rises by
typically at least an order of magnitude, usually within about a
second. This is followed by a decay, generally to the pre-burst X-ray
flux level, in a time interval between $\sim$10 s and about a minute; 
in rare cases bursts last longer. 

Two types of X-ray bursts can be distinguished (Hoffman et al. 1978),
called type I and II. The type I bursts show a distinct softening of
the X-ray spectrum during the decay of the burst. Their recurrence
times are generally of the order of hours and longer, but occasionally
as short as a few minutes. The spectral development in type II bursts
is much less pronounced than that in the type I bursts. 

\begin{figure} 
\centerline{
\psfig{figure=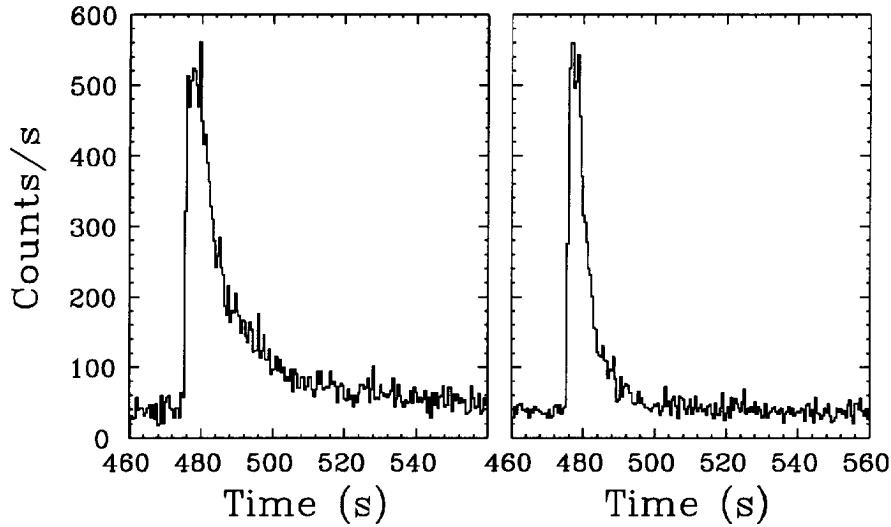}}
\caption{Type I  X-ray burst from 1702--42 as observed with Exosat in the
1.2 - 5.3 keV band (left) and the 5.3 - 19.0 keV band (right); the softening
of the X-ray burst spectrum is apparent as a longer tail in the low-energy
burst profile (courtesy T. Oosterbroek).} 
\end{figure}  

Until recently, the type II bursts had only been observed from the
Rapid Burster (Lewin et al. 1976); in a few other sources similar
events were possibly  observed. The time intervals between the type II
bursts from the Rapid Burster can be as short as $\sim$10 s, and as
long as one hour. They come in a characteristic pattern such that the
total energy in a burst is approximately proportional to the time
interval to the following burst: the Rapid Burster behaves like a
relaxation oscillator. This indicates that the type II bursts are the
result of an accretion instability. 

The sky distribution of X-ray bursters is strongly concentrated to the
center of the Galaxy. Thus, X-ray bursters are located at average
distances of $\sim$8 kpc (Reid 1993), and the total energy and maximum
luminosity in type I bursts are of the order of $10^{39}$ erg,  and
$10^{38}$ erg s$^{-1}$, respectively. All known  luminous LMXB in
globular clusters are X-ray bursters.

The ratio, $\alpha$, of the total energy emitted in the persistent
flux, to that emitted in bursts, is typically of the order $10^{2}$.
This is nicely accounted for by the thermonuclear-flash model of type
I X-ray bursts: after a sufficient amount of matter has accreted on
the neutron star surface, critical conditions may develop at the base
of the accreted layer, causing unstable helium burning. The sudden
release of nuclear energy gives rise to an X-ray burst. In this model
$\alpha$ is the ratio of  the gravitational binding energy to the
available thermonuclear energy per gram of accreted material (i.e., of
order $10^{2 \pm 0.5}$). Typical values for the rise time, decay time,
and recurrence time, and for the maximum luminosity and integrated
energy for type I X-ray bursts, are well reproduced by this model. 

Swank et al. (1977) found that for a particular burst they observed
the X-ray spectrum was best fit by a blackbody spectrum, with a
temperature that decreased during the decay of the burst. The
blackbody radius they found during burst decay  (assuming a distance
of 10 kpc) was $\sim$15 km. This indicates that the  radiation
observed during a type I X-ray burst originates directly from the
surface of the neutron star. This result forms the basis for attempts
to study the mass-radius relation of neutron stars from observations
of the spectral evolution during X-ray bursts. 

Several properties of type I X-ray bursts show a global correlation
with X-ray luminosity (i.e., mass accretion rate). Van Paradijs et al.
(1988) found that the burst duration $\tau$ and the above ratio
$\alpha$ are strongly correlated with the persistent X-ray luminosity 
$L_{\rm X}$, as measured (see Sect. 2.1.2) by the 
ratio $\gamma$ of persistent
X-ray flux to the peak flux of bursts with radius expansion (i.e.,
with peak luminosity equal to the Eddington limit, $L_{\rm Edd}$).
Above $L_{\rm X} \simeq 0.3 L_{\rm Edd}$ 
burst activity is extremely rare.
The decrease of the burst duration
with $\gamma$ indicates the decreasing importance of hydrogen in the
energetics of the thermonuclear flashes, as the persistent luminosity
increases. The strong increase of $\alpha$ with $\gamma$ they found,
implies that independent of the accretion rate, after a given waiting
time a source produces an X-ray burst with approximately the same
energy. This result is not accounted for by present
thermonuclear-flash models. Spallation of CNO nuclei in the accretion flow 
has been suggested as a possible origin of the behaviour of the $\alpha$ 
parameter (Bildsten, Salpeter \& Wasserman 1993).

A comprehensive review of X-ray bursts has been given by Lewin, Van
Paradijs \& Taam (1993, 1995). For a recent discussion of the physics of 
thermonuclear burning on the surfaces of accreting neutron stars, 
with emphasis on
the propagation of the burning, I refer to the contribution to this volume 
by L. Bildsten. 

\subsubsection{The Bursting Pulsar}

Several dozen X-ray pulsars are HMXB, but only four have so far been
found in LMXB. All known X-ray burst sources are LMXB. Until recently,
there was not a single X-ray binary known which emitted both bursts
and pulsations. This situation of mutual exclusion of bursts and
pulsations was recently terminated with the discovery (Kouveliotou et
al. 1996) of a transient source of hard X rays, GRO~J1744--28, which
between December 1995 and May 1996 emitted more than 3000 Type II
X-ray bursts. The persistent emission of this source shows 2.1 Hz
pulsations (Finger et al. 1996). This Bursting Pulsar is discussed in
the contributions to this Volume by C. Kouveliotou, and by L. 
Bildsten.

\subsubsection{Fast non-periodic variability; source states}   

Pulsations and bursts are easily identifiable phenomena that allow for
immediate interpretations in terms of neutron star properties (mass,
radius, magnetic field). Irregular variability, which has a less
immediate  diagnostic value, has been observed on many time scales
from early in the history of X-ray astronomy on (Lewin et al. 1968).
The early developments have been summarized by Bradt, Kelley \& Petro
(1982).

Our current view on fast irregular variability developed from the
discovery of intensity-dependent quasi-periodic oscillations in GX 5-1
(Van der Klis et al. 1985) and other LMXB, and the correlation of this
fast variability with the spectral properties of the X-ray source (see
Lewin et al. 1988 for a review of the early developments). This has
led to the recognition of two types of LMXB, with distinct correlated
temporal and spectral characteristics; after the shape of the tracks
that they follow in X-ray colour-colour diagrams (CCDs) these are
called the Z sources and the atoll sources (Hasinger \& Van der Klis
1989).  

Six Z sources are known. The Z-shaped track in the CCDs consist of the 
``horizontal'', ``normal'' and ``flaring'' branches (HB, NB, and FB,
respectively). All observed changes in the sources correspond to
continuous  movements along the track, jumps from one part to the
other do not occur. On each of the branches the rapid variability, as
expressed in the power density spectrum (PDS), has distinct
characteristics. On the HB the source intensity undergoes
quasi-periodic oscillations which in the PDS show up as a peak of finite
width, whose centroid frequency changes from $\sim 15$ Hz to $\sim 60$
Hz as the intensity increases and the source moves from the left side 
of the HB to the top of the NB. These HB oscillations are generally
interpreted as a modulation of a ``clumpy'' accretion flow at a
magnetospheric ``gate'', at a frequency equal to the spin frequency of
the neutron star as seen by a ``clump'' orbiting it with the Kepler
frequency (``beat frequency model'', Alpar \& Shaham 1985; Lamb et al. 1985). 
These HBO
disappear when the source reaches the upper part of the NB. The 
corresponding neutron star spin periods are in the millisecond range. In 
spite of major efforts, coherent pulsations have not yet been been detected 
in the persistent emission of LMXBs (see Vaughan et al. 1994). However, 
coherent 
oscillations with millisecond periods have recently been detected during 
some Type I X-ray bursts (see the contribution to this Volume 
by M. van der Klis).

Near the middle of the NB the PDS shows a different type of QPO, with
a near constant frequency of about 6 Hz (the same in all Z sources).
These NB oscillations are most likely the result of oscillations
of the optical depth (Stollman et al. 1987) of the accretion flow, 
caused by radiation feedback on the flow at a 
luminosity near the Eddington limit (Fortner et al. 1989; 
Alpar et al. 1992). A strong
argument for this model was provided by the observation in Cyg X-2 of
an  $\sim 180^{\circ}$ phase shift between oscillations above and
below  5 keV (Mitsuda \& Dotani 1989); 
this pivoting of the spectrum during the 6 Hz
oscillations reflects the variation in the Compton upscattering of
photons, as the scattering optical depth varies. As the Z source moves
from the NB to the FB the fequency of the 6 Hz oscillations increases,
and the width of the QPO peak in the PDS becomes much larger.

The optical and UV brightness of Z sources increases as the sources move from
the HB via the NB to the FB (Vrtilek et al. 1991; Augusteijn et al. 1992). 
This indicates that the mass accretion
rate increases along the Z track in that order. Since along the NB the
count rate decreases going from the HB to the FB this shows that count
rate can be anti-correlated with mass accretion rate.

The CCDs of atoll sources, many of which are burst sources, show a
``banana branch'', along which the source moves on a time scale of
hours, and ``islands'' where the source stays for much longer time
intervals. In the banana state the PDS contains only a power law
component, called the ``very-low-frequency noise''. In the island state
the PDS is dominated by a broad-band noise component, called
``high-frequency noise'', which has a power law shape at high
frequencies, but which flattens below a cut-off frequency $\nu_{\rm
co}$. Between the island state and the banana state the
mass accretion rate increases (Hasinger \& Van der Klis 1989).

Atoll sources  are on average much less luminous than Z sources; 
however, it is unlikely that the differences in their spectral and
temporal properties are only the result of a difference in accretion
rate. Hasinger \& Van der Klis (1989) suggested that the neutron star
magnetic fields in the atoll sources are systematically weaker than
those in the Z sources. 

Several broad noise components (e.g., ``very-low-frequency noise'',
``high-frequency noise'') in the power spectra of the Z sources  and
atoll sources have also been found in the power spectra of X-ray
pulsars and black-hole candidates. An example is the strong similarity
between the PDS of island-state atoll sources and those of accreting
black holes in the low (or ``hard'') state (see Sect. 3.3), including
the so-called Belloni-Hasinger (1990) effect, i.e., the  invariance of the
high-frequency part as $\nu_{\rm co}$ changes (Yoshida et al. 1993). A
proposal to unify the phenomenology of the power spectra of these
different types of accreting compact stars was made by Van der
Klis (1994). For a detailed description of aperiodic variability, and
its role in defining source states, I refer to Van der Klis (1995).

An important recent development is the detection with the Rossi X-ray
Timing Explorer of QPO with frequencies in the range 500-1200 Hz in
the PDS of Z and atoll sources. These are discussed in the
contribution to this Volume by M. van der Klis. 

\subsection{X-Ray Spectra}    

X-ray spectra played an important role in the development of the view
that there are different groups of compact X-ray sources; e.g.,
Tananbaum (1973) distinguished between ``X-ray binaries'' and ``Sco
X-1 type sources''. The former, which comprised systems like Cen X-3
with hard 1-10 keV X-ray spectra, are now known to be (mainly) HMXB
with strongly magnetized neutron stars: the X-ray binary pulsars. The
Sco X-1 type sources had relatively soft X-ray spectra (see Fig. 9);
they were suspected to belong to an old low-mass galactic population
(Salpeter 1973; see Section 1). For extensive discussions of the X-ray 
spectra of LMXB with neutron stars see White et al. (1984) and Mitsuda et 
al. (1984).

Ostriker (1977)  found that X-ray colour-colour diagrams provide an
efficient way to separate different groups of X-ray sources: he
suggested that, in addition to the above two types of X-ray binaries,
accreting black holes could be distinguished this way. The value of
these diagrams was shown by White \& Marshall (1984) who used them as
an efficient tool to distinguish accreting neutron stars with strong
and weak magnetic fields (by and large pulsating HMXB and
non-pulsating LMXB, respectively), and accreting black-hole
candidates. The spectra of the latter (both in HMXB and LMXB) showed a
strong low-energy excess whose strength appears to be related to the
mass accretion rate. 

\begin{figure}
\centerline{
\psfig{figure=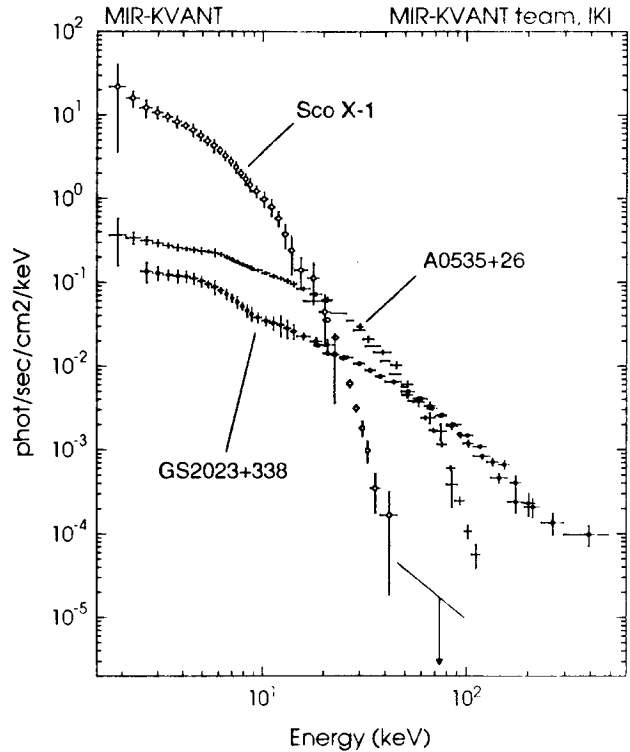}}
\caption{Spectra obtained with MIR-KVANT of three types of galactic X-ray 
binaries with different compact objects: Sco X-1 (weak-field neutron star), 
A~0535+26 (strong-field neutron star), and GS~2023+338 (a black hole). 
(From Gilfanov et al. 1995).}
\end{figure}

Fig. 9 shows typical examples of the X-ray spectra of accreting 
neutron stars with low and high magnetic fields, respectively, and of
an  accreting black hole. The spectra of luminous weak-field neutron
stars (e.g., Sco  X-1, X-ray burst sources) can be approximately
described by a  thermal-bremsstrahlung model, with $kT_{\rm TB} \sim
5$~keV. The  spectra of strong-field neutron stars (i.e., X-ray 
pulsars) are power laws (photon  indices $\sim 1$) with a high-energy
cut off at several tens of keV.

The difference in spectral hardness between HMXB and LMXB persists
into the hard X-ray range, up to $\sim10^{2}$ keV; it is remarkable
that the few LMXB which show pulsations (GX 1+4, Her X-1, 1627$-$673)
have X-ray spectra that are as hard as those of HMXB (almost all of
which are, likewise, pulsars). For a recent review of the hard X-ray
spectra of X-ray binaries see Gilfanov et al. (1995). Note that the X-ray 
spectra of the ``anomalous'' X-ray pulsars are very soft (see Stella et al.'s 
contribution to this Volume).

Above $\sim10$ keV the spectra of black-hole candidates often show a
hard power law component which in some cases can be detected up to
hundreds of keV. After the rejection of rapid variability as a
telltale sign of accreting black holes (Stella \etal 1985) this
high-energy tail became one of the criteria to select black-hole
candidates (see, e.g., Sunyaev  \etal 1991; see Sect. 3.2). 

The above simple picture of the spectra of X-ray binaries was enriched
by the work of the Granat/Sigma group (Barret et al. 1991, 1992;
Barret \& Vedrenne 1994), who found that  the
spectra of several low-luminosity X-ray burst sources (i.e., atoll
sources) have a power law high-energy tail extending to the $10^{2}$
keV range. These tails may be the high-energy extensions of the 1-20
keV power law spectra found earlier for some low-luminosity bursters
(Becker et al. 1977; see Barret \& Vedrenne 1994). Using the methods
for distance determinations described in Sect. 2.1.2, Van Paradijs \& Van
der Klis (1994) generalized this result by showing that there is a
general anti-correlation between the hardness of the $10-100$ keV
spectra of LMXB and their luminosity. A detailed discussion of the
hard X-ray spectra of X-ray bursters has been given by Barret et al.
(1996). These results strengthen the  similarity between accreting
black holes and atoll sources inferred from their power density
spectra. This subject is further discussed in Section
3.3. 

\subsection{Magnetic field decay ?}   

Many HMXB are X-ray pulsars, but X-ray pulsations occur only rarely in
LMXB. This suggests that the magnetic fields of the neutron stars in LMXB 
are generally much weaker than  in HMXB. An alternative
interpretation is that the magnetic and rotational axes of the neutron
stars in LMXB are aligned. However, the observation that the X-ray spectra
of LMXB are much softer than those of HMXB, except for the few LMXB
which do show pulsations (their X-ray spectra are as hard as those of
HMXB), strongly suggests that the division into hard and soft X-ray spectra
is related to a difference in the geometry of the accretion flow. For
neutron-star magnetic fields of the order of $10^{12}$\,G, and
sub-Eddington accretion rates, the accretion flow is dominated by the
magnetic field within a distance of $\sim 10^{3}$ km from the neutron
star; a large fraction  of the inflowing matter reaches the neutron star via
an accretion column on a relatively small area (near the polar caps). For
magnetic fields below $10^{9}$\,G one expects that the accreting material
is distributed over a larger fraction of the neutron star surface. 

It has long been known that not 
a single source shows both pulsations and type I bursts (note, however, 
that this may not be the case any more with the discovery with the Rossi 
XTE of coherent oscillations during some type I X-ray bursts; see the 
contributions to this Volume of M. van der Klis and L. Bildsten). The 
bursts from the Bursting Pulsar GRO\,J1744--28 are type II bursts 
(see Sect. 2.4.3).
Apparently, the presence of a strong magnetic field suppresses the
instability of the nuclear reactions that gives rise to bursts (as expected
from the very high rate of accretion locally inthe accretion column, 
see e.g., Joss \& Li 1980; for a recent 
discusion of the effect of magnetic fields on thermonuclear flashes, see 
Bildsten 1995). This
mutual exclusion of bursts and pulsations supports the idea that it is a
weaker magnetic field, and not only alignment of the field axis, which
distinguishes the neutron stars in LMXB from those in HMXB. 

There are two possible ways to understand this difference. In the first
place, the magnetic fields of the (generally old) neutron stars in LMXB may
be much weaker than those of the (young) neutron stars in HMXB, because
they have always been very weak. This difference might be related to a
difference in the formation mechanism of neutron stars in HMXB and LMXB,
$viz.$ via the normal evolution of a massive star, and via the
accretion-induced collapse of a white dwarf, respectively. 

An alternative
possibility is suggested by the observation that the neutron stars in HMXB
are all young objects, whereas those in LMXB are typically much older: the
magnetic fields of neutron stars decay. 
Until recently it was generally believed that all young neutron stars 
have strong dipolar magnetic fields $(B \sim 10^{12}$\,G), which decay 
spontaneously on a time scale of order $10^7$ years, to a bottom value of 
order $10^9$\,G. (The possibility that a substantial fraction of 
young neutron stars have magnetic fields much stronger than the canonical 
$10^{12}$\,G is discussed by D. Frail in this Volume). 
However, on the basis of a detailed analysis of the 
properties of the young radio pulsar population Bhattacharya et al. (1992) 
and Hartman et al. (1997) 
have shown that such spontaneous decay does not occur (with a lower limit  
to the exponential time scale for such decay of about $10^8$\,years). 

It has been suggested that decay of the magnetic field of a neutron star 
may be caused by the accretion process, either directly (e.g., by field 
burial, Romani 1990) or indirectly. A possibility that is currently 
investigated extensively (Srinivasan et al. 1990; Ding et al. 1993; 
Ruderman 1991a,b,c) is that before the phase of fully developed 
mass transfer the neutron star is spun down by the braking torques exerted 
by matter flowing in at a low rate from the secondary star (possibly induced 
by the pulsar wind) which is ejected at the rapidly rotating magnetosphere 
(``propeller effect''). The ``forced'' spin down of the neutron star (which 
occurs only if they are in a binary) corresponds to the outward motion of 
the quantized vortices (which carry the rotation of the superfluid core of 
the neutron star), whose interaction with the quantized magnetic flux tubes 
drags the latter along toward the crust, where Ohmic decay would lead to a 
decrease of the magnetic field (see the contribution of M. Ruderman to this 
Volume). 

Possible observational constraints on these considerations may be obtaind 
from the relation between the orbital period, $P_{\rm orb}$, and 
the neutron star spin period, $P_{\rm spin}$, observed for low-mass binary 
millisecond pulsars (LMBPs), the descendants of LMXBs (see Bhattacharya 
\& Van den Heuvel 1991, and Verbunt \& Van den Heuvel 1995, for reviews). 
One way to approach this 
is to assume some simple relation between the $B$ field 
of a neutron star (i.e., its dipolar component) and the amount, $\Delta M$, 
of matter accreted (see, e.g., Taam \& Van den Heuvel 1986). Combining this 
with a description of the evolution of LMXBs toward the LMBPs 
leads one to expect a relation between the orbital and spin 
periods of LMBPs, since $P_{\rm spin}$ is determined by $B$ and the 
mass transfer rate $\dot M$, and the latter depends on the orbital 
parameters (see, e.g., De Kool \& Van Paradijs 1987; Van den Heuvel \& 
Bitzaraki 1995; Li \& Wang 1997). For reasonable choices of the parameters 
in the $(B, \Delta M)$ relation these evolutionary scenarios can reproduce 
the observed $(P_{\rm orb},P_{\rm spin})$ relation for LMBPs quite 
well, but so does the spin-down induced field decay model (Jahan Min \& 
Bhattacharya 1994). 
It should be noted that accretion-induced magnetic-field decay 
does not necessarily imply a relation between $B$ and $\Delta M$. For 
instance, in the model of Konar \& Bhattacharya (1997) $B$ is related not 
to $\Delta M$ but to 
the rate, $\dot M$, of accretion.

A few LMBPs do not at all fit the $(P_{\rm orb}, P_{\rm spin})$  
relation; it has been argued that in these systems the neutron star was 
formed by the accretion induced collapse of a white dwarf (Van den Heuvel 
\& Bitzaraki 1995). However, as pointed out to me by Dr. R. Wijers, 
this argument depends on the acceptance of an assumed simple 
$(B,\Delta M)$ relation; the argument collapses if other parameters 
determine the evolution of the neutron star magnetic field (see also Wijers 
1997). 

The relation between accretion and a neutron star's 
magnetic field strength remains an unsolved issue, and its clarification is 
central to an understanding of the evolution of 
neutron stars in binaries.

\section{Black-hole X-ray binaries}

\subsection{Some background}

As discussed in Section 1, the first X-ray source which was shown to be a
member of a binary star, Cyg X-1, was a strong black-hole
candidate as well. In the words of the discoverers: ``The mass of the companion
probably being larger than about 2 M$_{\odot}$, it is inevitable that
we should also speculate that it might be a black hole.'' (Webster \&
Murdin 1972); ``This raises the distinct possibility that the secondary
is a black hole.'' (Bolton 1972).

In 1972 the radio source in the error box of Cyg X-1 showed a large 
brightness increase correlated with a major hardening of the 2-10 keV 
spectrum of Cyg X-1 (Tananbaum et al. 1972). We now know that the
spectral hardening is caused by the weakening of an `ultra-soft''
component in the X-ray spectrum, signalling a transition from a ``high''
(or ``soft'') state to a ``low'' (or ``hard'') state (see Section 3.3).

Following the discovery of the binary X-ray pulsar Cen X-3  (Schreier
et al. 1972) and many other similar systems, and  of X-ray bursters
(Grindlay et al. 1976;  Belian et al. 1976), research in X-ray 
binaries in the 1970's was dominated by systems in which the accreting 
compact object is a neutron star.  But research in black holes did not
disappear altogether. The discovery of strong rapid variability of the
X-ray flux of Cyg X-1  (Oda et al. 1971; see also Oda 1976, for a
review of early work on Cyg X-1) led to the idea that such variability
is a telltale sign  of an accreting black hole, which might be used
to distinguish them from  accreting neutron stars. On the basis of
this idea Cir X-1 was long  considered a black-hole candidate.
However, neutron star X-ray binaries can  also show rapid variability,
as was strikingly illustrated by the transient  V\,0332+53, which was
initially suggested as a possible black-hole system,  but later shown
to be an X-ray pulsar (Stella et al. 1985), whose pulse  amplitude
happened to be relatively weak compared to that of the red-noise
component in the PDS. Also Cir X-1 was shown to be  a neutron star
when it emitted type I X-ray bursts (Tennant et al. 1986).

Ostriker (1977) suggested that black-hole X-ray  binaries (BHXB) might
be distinguished  by the shape of their X-ray spectra. This idea was
put on a firm footing by White \& Marshall (1984)  who showed that in
an X-ray colour-colour diagram, derived from  the HEAO-1 A-2 sky
survey the two  sources, then known to contain black holes (Cyg X-1,
in its soft state, and LMC X-3) were located in the extreme
upper-left corner, i.e., their spectra were extremely soft.  A few
years later, McClintock  \& Remillard (1986) measured the mass function 
of the transient source A~0620--00 (which also had a very soft X-ray
spectrum during its outburst in 1975) after it had returned to
quiescence, to be $3.18 \pm 0.16$~M$_{\odot}$. This  immediately (see
below) showed that the compact star in this system is too massive to 
be a neutron star, and gave some confidence in the idea that X-ray
spectra may be an efficient way to select BHXBs.

In spite of the fact that some X-ray spectral characteristics of black
holes, and rapid variability are also seen in some neutron stars,
their combined presence, in particular in X-ray transients, has
remained strikingly  effective in singling out black holes.

As implied in the above discussion,  the main argument that the
compact object in a particular X-ray binary is a black hole, is  that
neutron star masses cannot exceed a certain maximum value. This
assumption rests on very general considerations, e.g., that sound
cannot travel faster than light, on the basis of which  Nauenberg \&
Chapline (1973) and Rhoades \& Ruffini (1974)  concluded that any
neutron star, independent of the equation of state (EOS) of
high-density matter, must have a mass $\lta 3$\,M$_{\odot}$.  Rotation
of the neutron star (ignored in the above analyses) does not increase
the mass limit by more than 20\% (Shapiro \& Teukolsky 1983). 
Detailed modelling of neutron stars, for a wide range of equations of
state, leads (see Fig. 10) to upper mass limits between $\sim
1.5$~M$_{\odot}$ (very soft EOS) and $\sim 2$~M$_{\odot}$ (very stiff
EOS) (see, e.g., Arnett \& Bowers 1977; Datta 1988; Cheng et al. 1993; Cook 
et al. 1994; Engvik et al. 1996; see also the contribution of N. 
Glendenning to this Volume). 

\begin{figure}
\centerline{
\psfig{figure=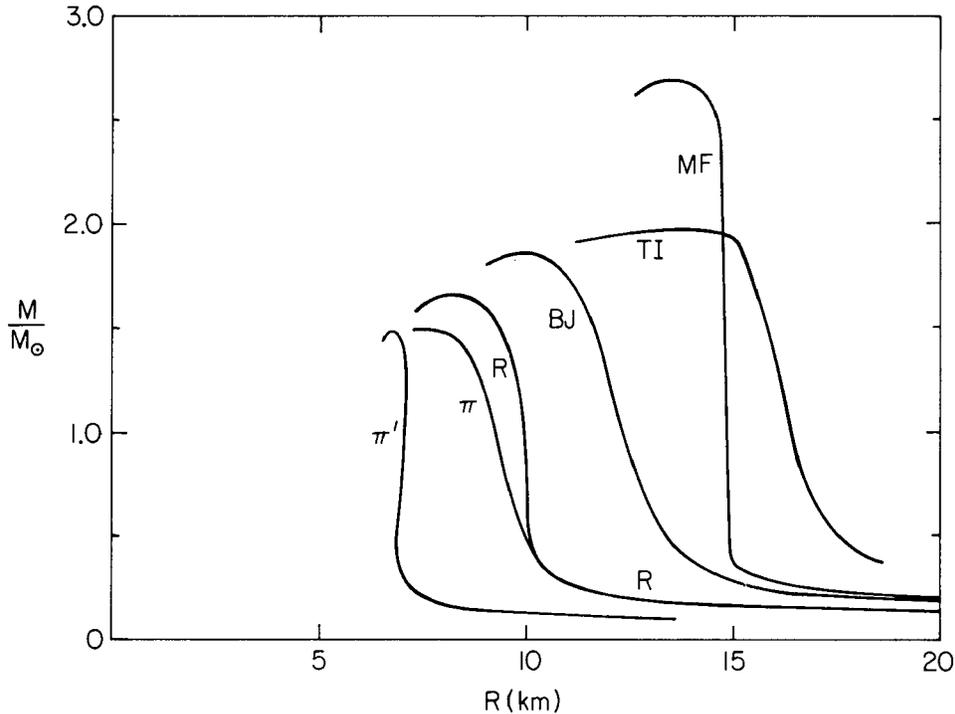}}
\caption{A selection of 
theoretical mass-radius relations for neutron stars, for several 
assumed equations of state (from Baym \& Pethick 1979).}
\end{figure}

The fact that compact objects with dynamical mass estimates exceeding
$\sim 3$~M$_{\odot}$ cannot be  neutron stars, is not equivalent to
their being  black holes, as defined by the particular space-time
structure described by Schwarzschild and Kerr metrics, which are
characterized, in particular, by the absence of a hard surface.  This
has led to the extensive use of the term ``black-hole candidate'' for
these objects. Of course, detection of X-ray pulsations or X-ray
bursts immediately disqualifies a compact star as a black hole, but
positive evidence for the absence of a hard surface has been very hard
to obtain. This should not come as a surprise, since a nominal
($M=1.4$\,M$_{\odot}$, $R=10$~km) neutron star is just 2.5 times
larger than its Schwarzschild radius, and one may expect the accretion
flow to be very similar to that of a black hole of comparable mass.
The energy release at the neutron star surface, which is absent for a
black hole, might lead to observable  differences in spectra and
variability, but unless the origin of the spectra and variability of
X-ray binaries is much better understood than it is nowadays, 
the conclusion that a black hole
has been found on the basis of such phenomena must be considered weak
at best. This difficulty is illustrated by the fact that the spectral
and variability characteristics of atoll sources are very similar to
those of black holes in their low state (see Sect. 3.3).

\begin{table} 
\caption{Black-Hole Binary Candidates from Radial-Velocity 
Measurements\hspace{2cm}}
\begin{center}
\begin{tabular}{lcccccc}
\hline \\
Name & Nova & $P_{\rm orb}$ & $f(M)$ & $i$ & $M_{\rm X}$ & Ref. \\ 
     &      & (days) &(M$_{\odot}$) & ($^{\circ}$) & (M$_{\odot}$) \\ \hline
{\it HMXB }\\
Cyg X-1      & No  & 5.6 & 0.25 & 28-38 &
$16 \pm 5$ & [1-3] \\
LMC X-3      & No  & 1.7 & 2.3 & 64-70 &
3.5-10 & [4,5] \\
LMC X-1      & No  & 4.2 & 0.144 & 40-63   &
 4-10 & [6] \\
\\
{\it LMXB } \\
A~0620--00    & Yes & 0.32 & 2.70 & 66.5-73.5 &
 3.3-4.2 & [7,8] \\
  &  &  &  &  31-54  & $14\pm7$  &  [9] \\
GS~2023+338  & Yes & 6.47 & 6.08 & 52-60 &
 10-15 & [10-12] \\
GS/GRS~1124--68   & Yes & 0.43 & 3.1 & 55-65  &
 4.5-7.5 & [13] \\
GRO~J0422+32     & Yes & 0.21 & 1.2 & 45-51   &
 3.2-3.9  & [14,15] \\
 &  &  &  &  13-31 & $>9$ & [16] \\
GRO~J1655--40 & Yes & 2.60 & 3.16 &  63-71 &
$7 \pm 0.7$ & [17-19] \\
GS~2000+25   & Yes & 0.35 & 5.0 & 43-85 &
4.8-14.4 & [20-23] \\ 
H~1705--25 & Yes & 0.52  & 4.0 & 60-80 & 
3.5-8.5  &  [24] \\ \hline

\end{tabular}
\end{center}
\footnotesize
[1] Webster \& Murdin (1972);
[2] Bolton (1972);
[3] Gies \& Bolton (1986)
[4] Cowley et al. (1983);
[5] Kuiper et al. (1988)
[6] Hutchings et al. (1987);
[7] McClintock \& Remillard (1986);
[8] Marsh et al. (1994);
[9] Shahbaz et al. (1994a);
[10] Casares et al. (1992);
[11] Casares \& Charles (1994);
[12] Shahbaz et al. (1994b);
[13] Remillard, McClintock \& Bailyn (1992);
[14] Filippenko, Matheson \& Ho (1995);
[15] Casares et. al. (1995a);
[16] Beekman et al. 1997; 
[17] Bailyn et al. (1995);
[18] Van der Hooft et al. (1997);
[19] Orosz \& Bailyn (1997);
[20] Casares, Charles \& Marsh (1995b);
[21] Filippenko, Matheson \& Barth (1995);
[22] Beekman et al.  (1996);
[23] Callanan et al. (1996);
[24] Remillard et al. (1996).
\end{table}

\normalsize

One way to infer the absence of a hard surface would be to show that
at some distance from the compact object matter flows inward  which
does not give rise, by a large margin, to the emission of X rays from
the compact object at the expected rate. Such evidence for the absence
of a hard surface has recently been presented by Narayan et al. (1996,
1997) from a comparative study of the X-ray properties of quiescent 
SXTs with black holes and neutron stars, respectively, and
by Belloni et al. (1997) for the microquasar GRS\,1915+105 during an
X-ray outburst.

Currently, ten X-ray binaries are known to contain black holes on the
basis  of a dynamical mass determination; seven of these are transient
low-mass  X-ray binaries.  Another 17 systems are suspected to be BHXB 
on the basis of their X-ray spectra (see
Tables 2 and 3, adapted from White \& Van Paradijs  1996).  The total number
of transient BHXB in the Galaxy is  $\sim 10^3$
(Tanaka \& Lewin 1995; White \& Van Paradijs 1996).

\begin{table} 
\caption{Black Hole Binary Candidates with Late-Type Companions\hspace{2cm}} 
\begin{center}
\begin{tabular}{lccc}
\hline \\
Name                    & Type & Signature$^a$   & Ref.  \\ \hline
GRO\,J0422+32 (V518 Per) & Nova & UH         & [1-3]   \\
A\,0620--00 (V616 Mon)   & Nova & US         & [4-6] \\
GRS\,1009--45 (XN Vel 93)& Nova & UH         & [7,8]   \\
GRS\,1124--68 (GU Mus)   & Nova & US/UH      & [9,10] \\
GS\,1354--645 (Cen X-2)  & Nova & US/UH      & [11]   \\
A\,1524--62 (TrA X-1)    & Nova & US/UH      & [12-14]   \\
4U\,1543--47             & Nova & US/UH      & [15-18]   \\
4U\,1630--47 (Nor X-1)   & Nova & US/UH      & [19-21]  \\
GRO\,J1655--40 (XN Sco 94)& Nova & US/UH/J   & [22-24]  \\
GX\,339--4 (4U1658--48)  & Variable & US/UH  & [25]     \\
H\,1705--25 (V2107 Oph)  & Nova & US/UH      & [26,27] \\
GRO\,J1719--24 (GRS\,1716--249) & Nova & UH  & [28-30] \\
KS\,1730--312            & Nova & UH         & [31]  \\
GRS\,1737--31            & Nova & UH         & [32] \\
1E\,1740.7--2942         & Variable & UH/J   & [33,34] \\
H\,1743--32              & Nova & US/UH      & [35-38] \\
SLX\,1746--331            & Nova & US        & [39]  \\ 
4U\,1755--338            & Persistent & US   & [40] \\
GRS\,1758--258           & Variable & UH/J   & [41,42] \\
GS\,1826--24$^b$         &  Variable & UH    & [43]  \\
EXO\,1846--031           & Nova & US/UH      & [44]  \\
GRS\,1915+105            & Variable & UH?/J  & [45-48] \\
4U\,1957+11              & Persistent & US   & [49] \\
GS\,2000+25 (QZ Vul)     & Nova & US/UH      & [50]  \\
GS\,2023+33 (V404 Cyg)   & Nova & UH         & [51,52] \\ \hline
\end{tabular}
\end{center}

\footnotesize
\noindent $^a$ UH: power law spectral component; US: ultrasoft 
spectral component; J: jets. 

\noindent $^b$ Recently, X-ray bursts were 
detected from this source with BeppoSAX (H. Muller, private communication). 

\noindent References: [1] Paciesas et al. 1992;
[2] Shrader et al. 1994; 
[3] Callanan et al. 1994; 
[4] Elvis et al. 1975; 
[5] Eachus, Wright \& Liller 1976; 
[6] Whelan et al. 1977; 
[7] Lapshov et al. 1993;
[8] Shahbaz et al. 1996; 
[9] Brandt et al. 1992;
[10] Kitamoto et al. 1992; 
[11] Kitamoto et al. 1990;
[12] Kaluzienski et al. 1975; 
[13] Barret, D. et al. 1995; 
[14] Barret, D. et al. 1992;
[15] Matilsky et al. 1972;
[16] Kitamoto et al. 1984;
[17] Harmon et al. 1992;
[18] Pedersen et al. 1983; 
[19] Jones et al. 1976;
[20] Priedhorsky \& Holt 1987;
[21] Parmar et al. 1995; 
[22] Harmon et al. 1995; 
[23] Bailyn et al. 1995;
[24] Van der Hooft et al. 1997; 
[25] Makishima et al. 1986; 
[26] Watson et al. 1977; 
[27] Griffiths et al. 1977; 
[28] Van der Hooft et al. 1996; 
[29] Ballet et al. 1993; 
[30] Harmon et al. 1993; 
[31] Sunyaev et al. 1994; 
[32] Cui et al. 1997; 
[33] Skinner et al. 1995; 
[34] Cook et al. 1995; 
[35] Kaluzienski \& Holt 1977; 
[36] Doxsey et al. 1977; 
[37] White et al. 1984; 
[38] Wood et al. 1984; 
[39] Skinner et al. 1990; 
[40] Church \& Balucienska-Church 1997; 
[41] Chen et al. 1994; 
[42] Mereghetti et al. 1996; 
[43] Makino 1988b; 
[44] Parmar et al. 1993;
[45] Castro-Tirado et al. 1994; 
[46] Belloni et al. 1997; 
[47] Morgan et al. 1997; 
[48] Taam et al. 1997; 
[49] Ricci et al. 1995; 
[50] Tsunemi et al. 1989; 
[51] Kitamoto et al. 1989; 
[52] Wagner et al. 1991.
\end{table}

\normalsize

\subsection{X-ray spectra of black-hole X-ray binaries}

In the X-ray spectra of BHXBs one finds two
components, whose  relative strengths can vary by a large factor (for 
recent reviews see Gilfanov et al. 1995; Tanaka \& Lewin 1995; Tanaka \& 
Shibazaki 1996; see Fig. 11).  
One is a power law, with  photon index in the range
$\sim 1.5$ to $\sim 2.5$, which dominates the  high-energy part ($\gta
10$~keV) of the spectrum, and is occasionally detected up to energies of
hundreds of keV. The other component  is generally limited to
photon energies below 10 keV, and is called the ``ultra-soft'' 
component. It is roughly described by a Planck function, with $kT_{\rm
bb}  < 1$~keV. 

\subsubsection{The ultra-soft component}

The ultra-soft component is generally interpreted as the emission from
an  optically thick, geometrically thin, accretion disk. In the limit
that relativistic effects are ignored, the temperature distribution
$T(r)$  for a standard  accretion disk is approximately given by 

\begin{equation}
T^{4}(r) = 3 G M_{\rm X} \dot M_{\rm d} / 8\pi\sigma r^3
\end{equation}

\noindent where $r$ is the radial distance from the center, 
$M_{\rm X}$  is the mass of the accreting object, and $\dot M_{\rm d}$  is the
mass transfer rate  through the optically thick disk (see, e.g., Frank, 
King \& Rainer 1992). Note that $\dot M_{\rm d}$  is not  
necessarily equal to the total mass accretion
rate, as part of the flow may  pass through a geometrically thick very
hot advective flow (Rees et al.  1982; Narayan 1996), or leave  the
system after moving inward through the disk (e.g., in a jet ejected
from the near vicinity of the compact star).

\begin{figure} 
\centerline{
\psfig{figure=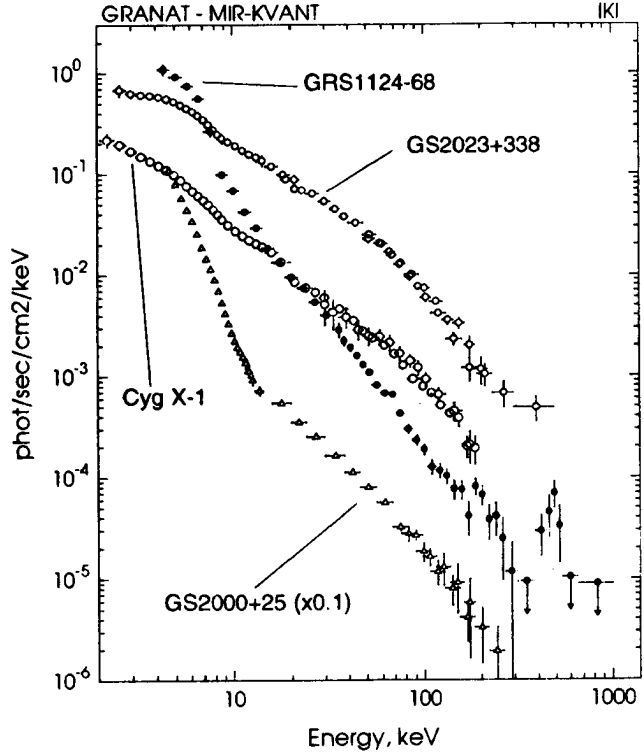}}
\caption{X-ray spectra of several
black-hole X-ray binaries, showing  various combinations of ultra-soft
and power law components (from Gilfanov  et al. 1995).} 
\end{figure}

Mitsuda et al. (1984) assumed that the local emission from the disk is 
Planckian, and derived the following expression (``multi-color disk
spectrum'', MCD) for the flux observed from the disk:

\begin{equation}
f(E) = {{8\pi R^2_{\rm in}~{\rm cos}~i~T^{8/3}_{\rm in}}\over{3d^2}} 
\int_{T_{\rm out}}^{T_{\rm in}} T^{-11/3}~B(E,T)~dT
\end{equation}

\noindent Here $i$ is the inclination angle of the disk, $R_{\rm in}$
is  the inner radius of the disk, $d$ is the source distance, $B(E,T)$
is the  Planck function, and $T_{\rm in}$ and $T_{\rm out}$ are the
disk  temperatures at the inner and outer disk radii, respectively. 
Usually, one sets $T_{\rm out} =0$. Note that the
disk is assumed to  be flat; therefore, at very high
inclination angles the model may  not be applicable, e.g., because of
self-occultation of the disk. In this model, if general-relativistic
effects are ignored,  the total disk luminosity $L_{\rm d}$ is given
by (Makishima et al. 1986)  $L_{\rm d} = 4\pi R_{\rm in}^2 \sigma
T_{\rm in}^4$, i.e., formally the  expression is the same as for a
spherical uniform blackbody emitter with  radius $R_{\rm in}$ and
temperature $T_{\rm in}$, although all disk emission  originates
outside $R_{\rm in}$.

The parameter $T_{\rm in}$ is determined from a fit of the MCD model to 
the shape of the
observed  spectrum; $R_{\rm in}$ is a factor in the normalization of
the fit, and  can be obtained if the distance and inclination angle 
are known. Together, $R_{\rm in}$ and $T_{\rm in}$ determine the mass
flow  rate through the disk, according to

\begin{equation}
\dot {M}_{\rm d} = 8 \pi \sigma R_{\rm in}^3 T_{\rm in}^4 /(3 GM_{\rm X})
\end{equation}

Usually, it is assumed that the inner disk radius is located at three
times the  Schwarzschild radius, inside of which stable orbits around
a non-rotating  black hole do not exist, i.e., $R_{\rm in} = 6 GM_{\rm
X}/c^2$. In that case (no relativistic effects included) 

\begin{equation}
T_{\rm in} = (1.58~ {\rm keV})~(M/1.4~{\rm M}_{\odot})^{-1/2}~
({\dot M}/10^{18}\,{\rm g/s})^{1/4} 
\end{equation}

Spectral fits with this model have been made to X-ray spectra obtained
with  {\it Ginga} throughout the outbursts of several transient BHXBs. 
Remarkably, $R_{\rm in}$ remained constant, while the disk luminosity
changed by more than an order of magnitude (see Tanaka \& Lewin 1995).
The values of $R_{\rm in}$ obtained from the fits are consistent  with
three times the Schwarzschild radius 
for stellar-mass black holes. This has given
some confidence in the applicability of the  model. 

Several factors complicate the interpretation of results obtained with
this simple model, the most important of which are:  (i) deviations of
the local disk emission from a Planck function; (ii) relativistic
effects. 

At the high temperatures encountered in the inner disk region the
opacity is dominated by electron scattering. This gives rise to a
hardening of the spectrum, which was first studied in the context of
hot neutron star atmospheres producing type I X-ray bursts (see, e.g.,
Van Paradijs 1982; London et al. 1984; Ebisuzaki 1987). To a reasonable
degree of approximation these burst spectra can be described by a
Planck function at a ``colour temperature'' $T_{\rm c}$ which exceeds the
effective temperature (as defined in the usual way by the total flux)
by a ``hardening factor'' $f = T_{\rm c}/T_{\rm eff}$, i.e., the flux,
$F_\nu$ is given by:

\begin{equation}
F_\nu = f^{-4}~B_\nu(fT_{\rm eff})
\end{equation}

Typical values of $f$ for X-ray burst spectra are near 1.5. Accretion
disks differ from hot neutron star atmospheres with respect to
atmospheric parameters, and because the latter passively transfer 
flux presented from below, whereas in accretion disk atmospheres
energy may be  dissipated in situ. Shimura \& Takahara (1995) made
numerical calculations of the spectra locally emitted in accretion
disks around black holes; their calculations included relativistic
effects for a non-rotating black hole (see below). They find that the
hardening factor $f$, as observed locally at the disk surface, varies
between $\sim 1.7$ and $\sim 1.9$, depending on the mass accretion
rate and the black-hole mass. The shapes of their integrated spectra
do not differ much from those of MCD spectra (of course, the fit
parameter $T_{\rm in}$ does not necessarily correspond to the inner disk
temperature in the numerical model).

Relativistic effects on disk spectra were discussed by Hanawa (1989)
based on earlier work by Page \& Thorne (1974): gravitational
redshift changes photon  energy and rate of arrival of photons as
measured by the observer; rotational velocities in the disk and 
bending of photon paths give rise to inclination dependent effects
(which ultimately requires folding in of the limb darkening of the
disk emission).
 
Numerical calculations of disk spectra, including relativistic effects
for a (non-rotating) Schwarzschild black hole, by Ebisawa et al.
(1993) show that these theoretical spectra (in which the detector 
response of Ginga was incorporated) cannot easily be distinguished
from the MCD functions, particularly not when they occur in
combination with another spectral component. Fitting MCD functions to
the numerically calculated spectra, Ebisawa et al. found 
the following expression for the fitted $T_{\rm in}$ parameters: 

\begin{equation}
T_{\rm in} = (1.47\,{\rm keV})~(M/1.4~{\rm M}_{\odot})^{-1/2}~
({\dot M}/10^{18}{\rm g/s})^{1/4}~(f/1.5)~g_2(i)
\end{equation}

\noindent (here $f$ is the spectral hardening factor, and $g_2(i)$ varies
between 0.76 and 1.0). This is quite similar to the ``Newtonian'' values
of $T_{\rm in}$ (see Eq. 4). It therefore appears that the continuum
disk spectra are not affected in a major way by relativistic effects.
Conversely, it appears  difficult to obtain information on these
effects from studies of the disk continuum. 

Compared to continuum emissions spectral lines 
have superior diagnostic value, since radial-distance
(redshift) information is partially preserved through the energy
resolution of the data. Gravitational redshift and the relativistic Doppler 
effect can produce very asymmetric line profiles (see, e.g., Laor 1991), 
which have been observed in the spectra of some AGNs, but not, so far, 
in the X-ray spectra of X-ray binaries (Tanaka et al. 1995; Fabian et al. 
1995). 

As shown by Thorne (1974), rotation is likely to be important for accreting
black holes. It strongly affects the radius of the innermost stable
orbit,  which may be as small as the radius of the horizon (for
maximum prograde disk rotation), or as large as 9 times that (maximum
retrograde disk rotation). S.N. Zhang et al. (1997c) showed that information
about the rotation of a black hole of known mass may be obtained from
a comparison of the inner disk radius $R_{\rm in}$ (derived from
spectral fits) with the  expected (Schwarzschild) value, and pointed
out further possible observational consequences of such rotation.

\subsubsection{The power law component}

X-ray spectra of BHXBs often contain a very hard power
law component, which in some cases has been observed up to the MeV
range (see Fig. 11). This hard X-ray component is generally
interpreted as  the result of Compton up-scattering of low-energy
photons in a very hot  medium, generally associated with a disk
corona, or a geometrically thick  inner disk (Sunyaev \& Titarchuk
1980).  Approximating this spectral component as a power law with an 
exponential cutoff at high energies, the photon index, $\Gamma$,  of
the power law is given by  $\Gamma \approx -{1/2} + \sqrt{{9/4} +
\pi^2/3y}$, where $y$ is the  Compton parameter $y = 4kT\tau^2/m_{\rm
e}c^2$ ($T$ and $\tau$ are  the temperature and the scattering optical
depth  of the hot electron gas). 

The nature of the very hot electron gas is not immediately obvious.
The  heating mechanism may be related to magnetic processes on the
disk surface,  analogous to coronal heating in late-type stars. The
hot scattering  medium may be a by-product of an advective flow in
which the ion  temperature is of order the virial temperature (Rees et
al. 1982; Narayan  1996); the electron temperature  is then determined
by a balance between heating due to Coulomb interactions  with the
ions, and cooling due to upscattering of low-energy photons.  It has
been  suggested that scattering may occur on a converging bulk flow in
the near  vicinity of a black hole (Blandford \& Payne 1981; Payne \&
Blandford  1981; Chakrabarthy \& Titarchuk 1995); the expected photon
index in  this case is $\sim 2.5$.

Although thermal Comptonization models have been succesful in
describing the hard X-ray spectra of BHXB in the $\sim 10^2$ keV
range, they do not easily account for the $\sim 1$ MeV emission
detected from Cyg X-1 and the transient GRO\,J0422+32 (Phlips et al.
1996; Van Dijk et al. 1995). Particle acceleration producing a
non-Maxwellian tail of the electron distribution may be necessary to
explain this emission (Skibo \& Dermer 1995; Li \& Miller 1997).

In case the Comptonization processes are not saturated one expects
that, on average, the higher the observed photon energy is, the more
scatterings the photon underwent in the hot medium; therefore,
higher-energy photons are expected to be delayed relative to
lower-energy photons, by an amount that depends on the size and
optical depth of the Comptonizing medium. Therefore, measurements of
energy-dependent time delays in X-ray flux variations may produce
constraints on the properties of the scattering medium (Sunyaev \&
Titarchuk 1980;  Stollman et al. 1987; Wijers et al. 1987; Bussard et
al. 1988; Miller 1995). For unsaturated Comptonization the time delay
between photons at energies $E_1$ and $E_2$ is expected to be
proportional to ln\,$(E_1/E_2)$ (Payne 1980; Sunyaev \& Titarchuk 1980). 

Delay measurements have been made for Cyg X-1 (Miyamoto et al. 1988,
1989; Miyamoti \& Kitamoto 1989; Crary et al. 1997) and 
several other BHXBs (Miyamoto et al. 1991, 1993). 
The energy dependence of the delays measured for Cyg X-1 are 
proportional to ln$(E_1/E_2)$ over  the range $\sim 5 - 100$ keV 
(Crary et al. 1997), supporting the Comptonization picture. However,
the time delays change with Fourier frequency $\nu$ roughly as
$\nu^{-0.8}$ (Crary et al. 1997), which is difficult to reconcile with
simple versions of the  Comptonization model, and may reflect
properties of the input photons before they are Comptonized. 

\subsection{Source States}

As discussed in Sect. 2.5,  our understanding of accreting weak-field 
neutron stars has much improved by the introduction of the concept  of
``source states'', defined by both temporal and spectral  properties of
these X-ray sources. This  concept has also proven fruitful in
studying BHXBs (see Van der Klis 1994, 1995), where
the simultaneous analysis of the spectra  and the fast variability has
led to the distinction of source states, as  follows (see Fig. 12). 

In the low state (LS) the ultra-soft spectral component is weak or 
absent, so the  X-ray spectrum is dominated by the hard power law
component. The PDS then shows  a strong broad-band noise component,
which at high frequencies is a power  law with slope $\sim -2$ with a
variable low-frequency cut off  $\nu_{\rm co}$; observed values for
$\nu_{\rm co}$ are in the range $\sim 0.02$ to $\sim 0.5$ Hz (Mendez \& Van 
der Klis 1997). 
Belloni  \& Hasinger (1990)
found that as $\nu_{\rm co}$ changes the  high-frequency part of the
PDS remains the same; thus the lower $\nu_{\rm co}$ is, the higher the
fractional variations in the X-ray flux.  From the long-term
monitoring with BATSE Crary et al. (1996)  found that the r.m.s.
fractional hard X-ray variability of Cyg X-1 is  correlated with the
photon index of the power law component in the X-ray  spectrum.

In the high state (HS) the X-ray spectrum shows the ultra-soft component, 
which in some cases completely dominates the emission. The amplitude of 
X-ray flux variations is then much smaller than in the low state; in the 
PDS they are represented by a weak power law. 

\begin{figure}
\centerline{
\psfig{figure=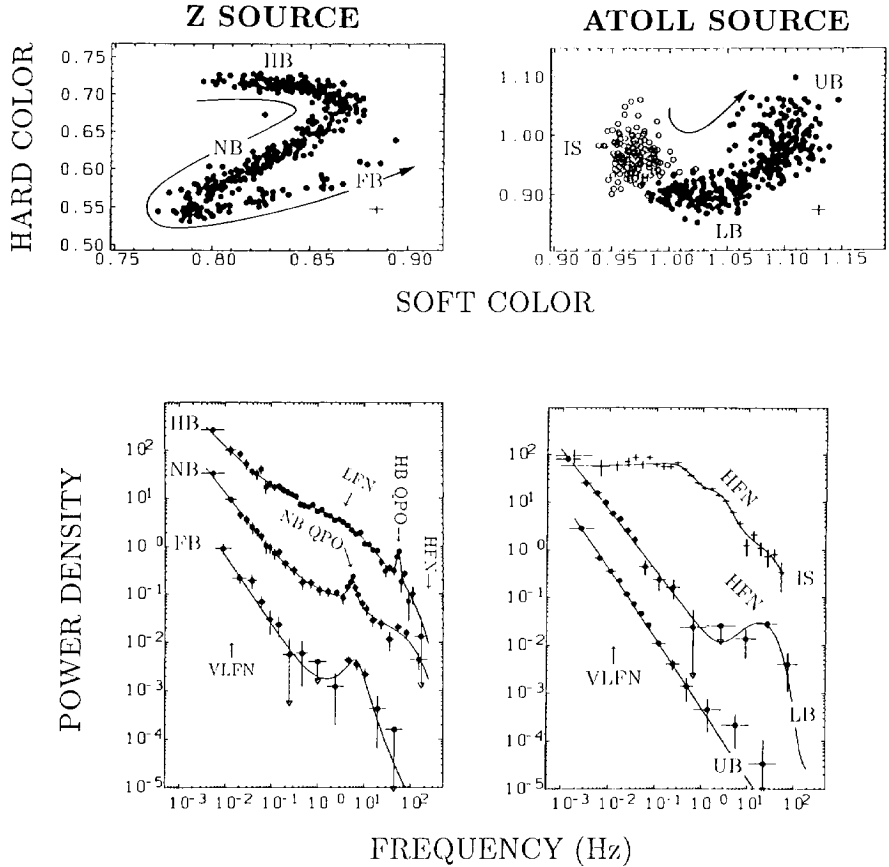}}
\caption{X-ray colour-colour diagrams and power density spectra typical of 
Z sources and atoll sources (Van der Klis 1995).}
\end{figure}

In the very high state (VHS) the X-ray spectrum contains the strong
ultra-soft  component, and a relatively steep power law component at
high energies.  The PDS shows, in addition to a broad-band noise
component (perhaps similar to that seen in the LS, Mendez \& Van der Klis 
1997), strong QPO with frequencies of order 10 Hz and substantial
harmonic content.  In N Mus 1991 the VHS was observed at the peak of
the outburst; afterwards  this source went into the HS which, in turn,
during the decay of the  outburst was followed by the LS. On the basis
of this ordering Van  der Klis (1994) concluded that the
sequence LS $\rightarrow$ HS $\rightarrow$ VHS is one of 
increasing mass accretion rate.

\begin{figure}
\centerline{
\psfig{figure=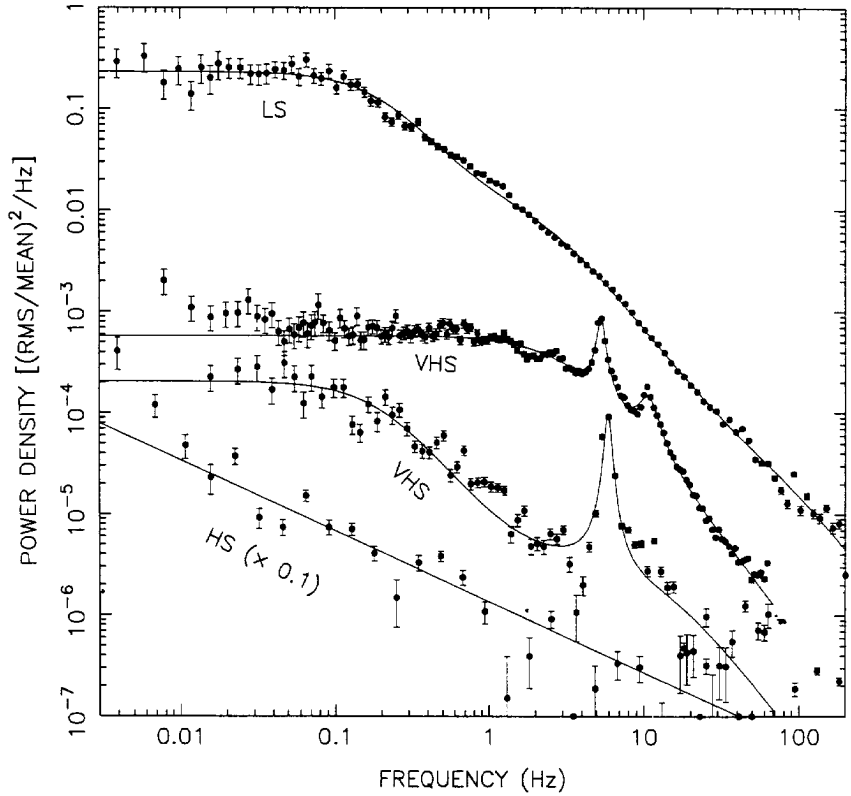}}
\caption{Power spectra from {\it Ginga} data of 
black-hole X-ray binaries in the 
low state (Cyg X-1), high and very-high states (GS~1124--68) (from 
Van der Klis 1995).}
\end{figure}

Recently, this pleasingly simple global picture of black-hole states
was  affected by the discovery that the PDS of N Mus 1991 showed VHS 
characteristics when the source moved from the HS to the LS (Belloni et al. 
1996a). This
possible  intermediate state has also been found in Cyg X-1 (Belloni
et al. 1996b) and GX\,339--4 (Mendez \& Van der Klis 1997).
 
Long-term monitoring of Cyg X-1 with BATSE and the All-Sky Monitor on
the Rossi XTE showed that major changes in the ultra-soft component
can occur on a time scale of order a day, whereas the power law
component changes much more slowly (S.N. Zhang et al. 1997a).
Contemporaneous radio monitoring during a soft-to-hard spectral
transisiton indicates a strong anti-correlation between the radio 
flux and the strength of the  ultra-soft component (S.N. Zhang et al.
1997b); this confirms the early results of Tananbaum et al. (1972),
and puts them on a much firmer observational basis.

Work by the Granat/Sigma group (see, e.g., Barret \& Vedrenne 1994)
showed  that at low luminosities the X-ray spectra of some LMXB with
neutron stars  are relatively hard power laws. Van Paradijs \& Van der
Klis (1994)  showed that there is  a general anti-correlation between
spectral hardness in the 13-80 keV range  and X-ray luminosity, and
that the X-ray spectra of NS-LMXB with  $L_{\rm X} \sim 10^{-2}\,{\rm
L}_{\rm Edd}$ (likely island state atoll sources) are as hard as those
of  black-hole binaries (see also the review of Gilfanov et al. 1995). 
Currently available data (see the comprehensive summary by Barret et
al. 1996) are consistent with the idea that only BHXB can show the
combination  of a hard power law X-ray  spectrum and a high X-ray
luminosity ($\gta 10^{37}$ erg/s).

The LS power spectra of accreting black holes are strikingly similar 
to those of atoll 
sources in the island state (see Fig. 14), not only with respect to the 
shape of the broad-band noise component, but also with respect to the 
invariance of its high-frequency part as the low-frequency cut off changes 
(Belloni-Hasinger effect). Together with the spectral similarities this 
suggests that at low accretion rates the dominant factors in the 
emission processes near black holes in the LS and near weakly magnetized 
neutron stars in the island state are the same; in particular, the presence 
of a hard surface in the case of accreting neutron stars,  
and possibly of a significant magnetic field, do not 
appear to have much effect on the X-ray spectral and temporal properties.

Another example of strong similarity between accreting weak-field 
neutron stars and black holes is provided by Cir X-1, whose  PDS, when
it is very bright, is very similar to those  of BHXBs in the VHS (see
Van der Klis 1995), in  particular by the presence of a high-frequency
broad-band noise component  and QPO.  This is  consistent with the
idea (Hasinger \& Van der Klis 1989)  that Cir X-1 is an atoll source
(i.e., the  neutron star magnetic field is very weak) accreting near
the Eddington  limit. The black-hole power spectra in the VHS (but not
that of Cir X-1) show higher harmonics of  these QPO; this may be a
distinguishing property of these objects. 

\begin{figure}
\centerline{
\psfig{figure=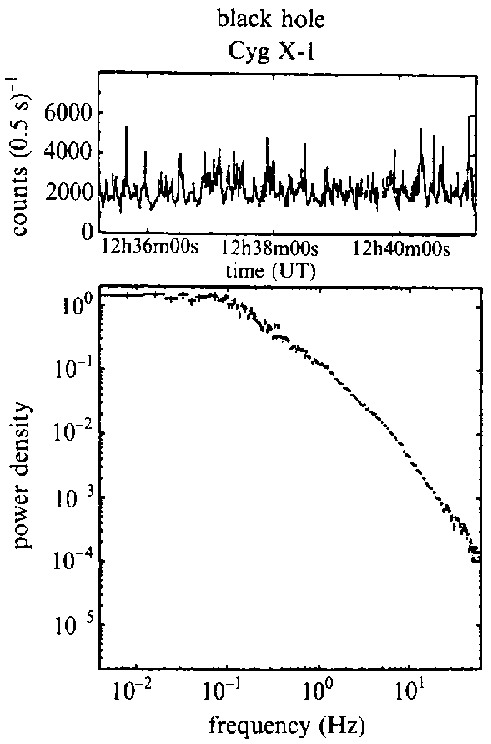}
\psfig{figure=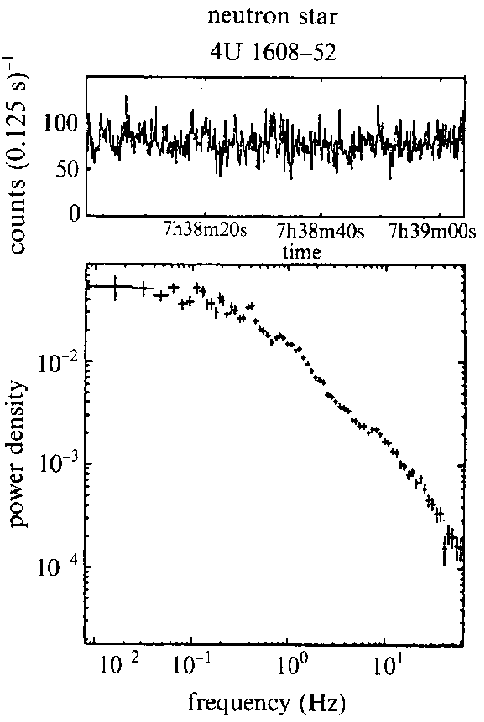}
}
\caption{Light curves and power spectra of a black hole (Cyg X-1) in the 
low state and an atoll source (4U~1608--52) in the island state (from Van 
der Klis 1995).}
\end{figure}

\subsection{Soft X-ray Transients}

A remarkably high fraction of the currently known BHXB are LMXB
transients (see Tables 2 and 3). Since these transients were first
studied in the 2-10 keV range, where the ultra-soft component can be
seen, they are often referred to as soft X-ray transients (SXT), in
spite of their often very hard power law spectra at higher energies.
SXT are characterized by sudden relatively brief increases in X-ray
luminosity (outbursts), separated by long periods of very low X-ray
emission (quiescence). The observed intervals between outbursts range
from about a year to 60 years, and it is highly likely that all SXT
are recurrent. During the outburst the optical emission from SXT
increases by a large factor as well, and has properties very similar
to that of persistently bright LMXB (see Sect. 2.1.2). 

>From the comprehensive summary of the properties of SXT by Chen et al.
(1997) it appears that typical outburst rise and decay  times are a
few days, and about a month, respectively. However, one cannot be but
struck by the variety in the observed outburst histories (see Fig.
15). It is not always  clear, though, in how far this reflect
variations in mass accretion rate, rather than major changes in the
X-ray spectra. 

\begin{figure}
\centerline{
\psfig{figure=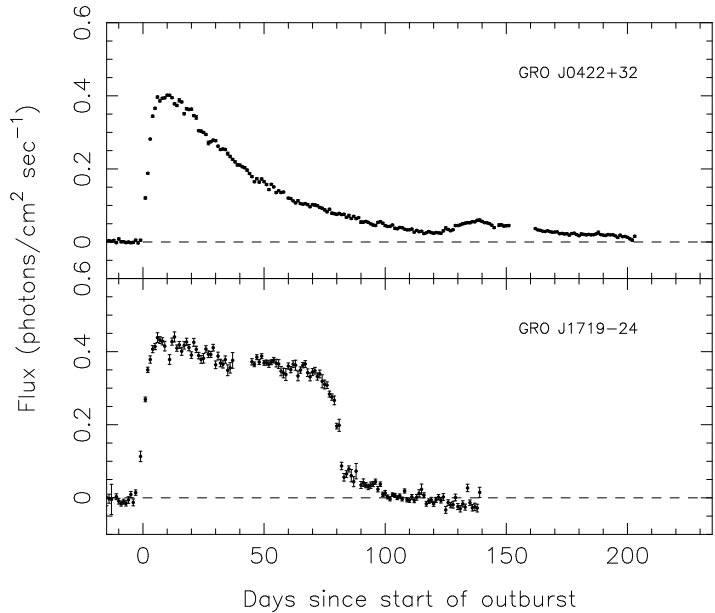,width=11.5cm,angle=-90}}
\caption{Outburst light curves of the BHXTs GRO~J0422+32 (XN Per 1992) and 
GRO~J1719--24 (XN Oph 1993). Courtesy F. van der Hooft.}
\end{figure}

The outbursts are caused by the sudden increase of the rate of accretion
onto the compact object. This increase is most likely caused by a disk
instability, analogous to that operating in dwarf novae, which occurs
if the disk is sufficiently cool, i.e., the mass transfer rate, $\dot
M$, is below a critical value (which depends on the orbital period).
These critical values for SXTs are much lower than those for dwarf
novae, because at a given value of $\dot M$ X-ray heating can keep the
disk hotter in LMXB than in CVs, which stabilizes them (Van Paradijs
1996; King et al. 1996; see Van Paradijs \& Verbunt 1984, for 
a comparison of the
properties of SXTs and dwarf novae).

During the outburst the optical emission is completely dominated by
reprocessing of X rays in the accretion disk, but in quiescence the
optical emission betrays the presence of the low-mass companion
(Thorstensen et al. 1978; Van Paradijs et al. 1980; McClintock \&
Remillard 1986). This allows the measurement of Doppler shifts
(McClintock \& Remillard 1986), which in so far seven cases has led to
dynamical evidence that the compact object in these SXTs is a black
hole (see Table 2).

During quiescence, the X-ray luminosity, $L_{\rm X}$, appears to
depend strongly on the nature of the compact star: for neutron star
systems it is of order $10^{33}$ erg\,s$^{-1}$ (Van Paradijs et al. 1987; 
Verbunt et al. 1994), for black holes
it can be as small as $\lta 10^{31}$ erg\,s$^{-1}$ (Narayan et al. 1996, 1997). 
On the basis of a detailed comparative study 
of the quiescent spectra of SXT with black holes and neutron stars, 
respectively, covering the optical to
X-ray ranges, Narayan et al. have argued that this
difference can only be understood if the flow in the inner disk of
quiescent SXTs is advective. At low mass transfer rates the
time scale for radiative cooling of the inner disk becomes longer than the
radial-flow time scale, and liberation of 
gravitational potential energy leads to an
increase of internal energy of matter in the disk. In the case of a
black hole most of this internal energy can 
flow through the horizon of the hole, but for neutron stars it
must emerge as it reaches  the neutron star surface. The detailed
spectral modelling by Narayan et al. (1996, 1997) has provided 
good evidence that the objects that we suspect are black holes on the
basis of their mass functions, do not have a hard surface as is
expected from general relativity. A similar argument has been put
forward by Belloni et al. (1997) for GRS\,1915+105 (at extremely high 
mass transfer rates).

An exciting result of the spectroscopic investigations of SXTs in
quiescence has been the discovery of very high abundances of Li in the
atmospheres of the  low-mass companions of the BHXTs 
A\,0620--00 (Marsh et al. 1994), V404\,Cyg (Martin et al. 1992), and N Mus
1991 (Casares et al. 1997), and the neutron star system 
Cen X-4 (Martin et al. 1994b). Since these
G-K type stars deplete Li on a short time scale, this observation
implies a source of recent Li production in these systems. Since Li is
not produced in ordinary nucleosynthesis, special environments are
required. Two possibilities have been suggested for this Li
production. 

\noindent (i) Podsiadlowski et al. (1995) proposed that SXTs are the
descendants of massive Thorne-\.Zytkow objects, i.e., red supergiants
with a neutron core which formed as a result of common-envelope
evolution. Around the neutron core is a very hot convective
nuclear-burning region; a high Li abundance may build up if
the residence time of matter in the lithium producing region is short
enough that most of the lithium formed there is transported outward to
cooler regions where it cannot be destroyed any more. In the later
phases of the Thorne-\.Zytkow object neutrino losses become the
dominant energy loss mechanism, leading to an effective reduction of
the Eddington limit. The resulting large rates of accretion onto the
neutron core may lead to the formation of a black hole. The secondary is 
formed from the collapsing envelope of the Thorne-\.Zytkow object. 

\noindent (ii) Martin et al. (1994a) have argued that Li is produced
in the SXT systems themselves, during the outbursts, by spallation of
CNO elements  (mainly by protons) as well as $\alpha \alpha$
reactions. This requires that at least a few percent of the accretion
energy must be converted to particle acceleration to energies
exceeding 10 MeV/nucleon (Martin et al. 1994b). An advective flow near the 
compact star may provide the environment in which the spallation occurs 
(Yi \& Narayan 1997).

The absence of certified black holes in persistent LMXB of course
reflects the severe difficulty, if not impossibility, of measuring the
orbital motion of the low-mass companion. However, this does not
explain the preponderance of black holes in optically identified SXTs.
To be transient requires the LMXB to have a very low mass transfer
rate (White et al. 1984; Van Paradijs 1996). The preponderance of
black holes in SXTs may then be partially explained by the larger
amount of orbital angular momentum, roughly by a factor 
$(M_{\rm BH}/M_{\rm NS})^{2/3}$, that has to be removed to produce the same
rate of mass transfer. King et al. (1996) have argued that the low
mass transfer rates imply that the secondaries of SXTs are  evolved.
This fits the observed very low secondary masses in Cen X-4 (Chevalier 
et al. 1989; McClintock \& Remillard 1990), and GRO\,J1744--28 (Finger et al.
1996). 

\subsection{Relativistic jets}

Mirabel \& Rodriguez (1994) found that the X-ray transient
GRS~1915+105  ejected radio emitting ``blobs'' in opposite directions,
with an angular  speed that appeared superluminal at the distance
$D=12.5$~kpc estimated  from 21 cm absorption line observations.
Using the expression  describing relativistic proper motion
($\mu_{1,2}$) of emitters ejected symmetrically at speed $V$ 

\begin{equation}
\mu_{1,2} = {{\beta~{\rm sin}~\theta} \over {1\pm \beta~{\rm cos}~\theta}}
{{c}\over{D}}
\end{equation}

\noindent they derived from $\mu_1 = 17.6$~mas per day, and 
$\mu_2 = 9.0$~mas per day that $\beta = V/c = 0.92$. 

Jets and two-sided ejection had been found before in the X-ray
binaries 
1E~1740.7--2942 (Mirabel et al. 1992), 
GRS~1758--258 (Rodriguez et al. 1993), 
SS~433 (see Margon 1984, for a review), 1E~0236+610 (Massi et al. 1993), 
and more recently in Cyg X-3 (Mioduszewski et al. 1997).
The nature of the compact objects in the last three  sources is 
unknown; the first two objects likely contain black holes on  the basis
of their hard power law X-ray spectra and high X-ray luminosity (see
Sect. 3.3). The connection between accreting black holes and
relativistic jets was  strengthened by the discovery (Hjellming \&
Rupen 1995) that also the X-ray  transient GRO~J1655--40 showed
superluminal expansion of double-sided  radio ``blobs'' ($\beta = 0.92$
for this system as well).  The mass function for this system (Bailyn
et al. 1995)  indicates that it contains a black hole (see Table 1). 

These exciting results provide a strong link between galactic black-hole X-ray 
binaries and active galactic nuclei (AGNs), a subset of which  eject
superluminal radio jets (see Antonucci 1993,  for a recent review of
AGNs),  and give strong support to the relativistic interpretation of 
superluminal motion in AGN. The  link is reinforced by the recent
finding of Sams  et al. (1996) that both BH X-ray binaries and AGN
with relativistic jets  follow one relation between the  size and
surface brightness of the jets and the accretion rate onto the  black
hole. 

The occurrence of jets appears to be independent of the nature of the 
accreting object. A recent review on the occurrence of jets, and the 
implications of this finding for the physics of jet formation has been 
given by Livio (1997).

\section{Mass determinations of Compact Stars in X-ray binaries}

\subsection{Neutron star masses and equation of state}

Apart from their crucial role in distinguishing black holes from
neutron stars, the importance of  measuring the masses of compact
stars in X-ray binaries is that they may provide constraints on the
properties of the high-density matter in the interior of neutron
stars. 

These properties are described by an equation of state (EOS), which
together with the Oppenheimer-Volkov equations allows one to calculate
models of the interior structure of neutron stars (see, e.g., Shapiro
\& Teukolsky 1983).  Since neutron stars can be considered to be
zero-temperature objects these models form a one-parameter sequence in
which mass, $M$, and radius, $R$, depend only on the central density.
For a given equation of state one thus has a unique mass-radius
relation.  Extensive calculations of neutron star models have been
made by Arnett \& Bowers (1977) and Datta (1988); for a detailed discussion 
I refer to the contribution of N. Glendenning to this Volume. 

Equations of state can be conveniently distinguished by the
compressibility of the neutron star matter; for very ``stiff'' and very
``soft'' EOS one finds that neutron stars have radii of $\sim 15$ km,
and $\sim 8$ km, respectively (see Fig. 10). Also, the maximum
possible neutron star mass depends on the EOS; it is $\sim
1.5$\,M$_{\odot}$ for very soft EOS, and up to $\sim 2.5$\,M$_{\odot}$
for the stiffest EOS. 

As will be discussed in more detail below, most neutron star masses
are consistent with a value close to 1.4\,M$_{\odot}$. From Fig. 10 it
appears that at this value masses do not allow one to draw conclusions
about the stiffness of the EOS of neutron star matter. For that, one
would need observed masses in excess of 1.6\,M$_{\odot}$, which would
exclude the softest EOS (note that stiff equations of state are not
excluded by low neutron star masses). Similarly, measurements of the
gravitational redshift, $z$, at the neutron star surface alone are not a
sensitive EOS discriminant, since both stiff and soft equations of
state allow $M/R$ ratios up to $\sim 0.2$\,M$_{\odot}$\,km$^{-1}$ (see
Fig. 10), corresponding to redshifts up to $\sim 0.6$.

Very accurate neutron star masses have been determined from a variety
of general-relativistic effects on the radio pulse arrival times of
double neutron star systems. These results will be briefly summarized
in Sect. 4.2.1. Neutron star masses have been determined for six 
HMXB pulsars from pulse arrival time measurements, in combination with 
radial-velocity observations of their massive companions (see Sect.
4.3). Masses have also been estimated for the low-mass 
binary radio pulsar PSR\,J1012+5307, whose companion is a white 
dwarf, and for the neutron stars in the
LMXBs Cyg X-2 (a Z source), Cen X-4 (an SXT) and 4U\,1626--67 (an X-ray pulsar).
These results are described in Sections 4.2.1, 4.3.3, and 4.3.4, 
respectively.

In addition to direct measurements
of mass and radius, a variety of other  ways to obtain observational
constraints on the EOS of neutron stars have been proposed. I will
limit myself here to just mentioning them. 

\noindent (i) Measurement of the
limiting spin period of neutron stars (Friedman et al. 1986; see also 
Glendenning's contribution to this Volume). 

\noindent (ii) The
cooling history of neutron stars (see Becker \& Tr\"umper's
contribution to this Volume). 

\noindent (iii) Measurement of the neutron star
magnetic field from the energy of a cyclotron line in the spectrum of
an X-ray pulsar, combined with an interpretation of its spin behaviour
in terms of an accretion torque model (Wassermann \& Shapiro 1983).

\noindent (iv) Glitches in the spin period, and its derivative, of radio pulsars
(Alpar, this Volume). 

\noindent (v) The neutrino light curve during a supernova
explosion (Loredo \& Lamb 1989). 

\noindent (vi) Measurements of kHz quasi-periodic 
oscillations in the X-ray intensity of LMXB, interpreted as Keplerian 
frequencies of orbits around neutron stars may provide constraints on the 
mass-radius relation of neutron stars by requiring that the neutron star 
and the innermost stable orbit around it, must fit inside the Kepler orbit 
(Kluzniak \& Wagoner 1985; Miller, Psaltis \& Lamb 1996; 
Kaaret \& Ford 1997; W. Zhang et al. 1997). A detailed discussion of these 
kHz QPO is given in the contribution to this Volume by M. van der Klis. 

\noindent (vii) Gravitational bending of X rays, as inferred 
from the X-ray pulse
profile for a radio pulsar for which the emission geometry is
determined from the radio pulse properties (Yancopoulos et al. 1994).

\noindent (viii) Measurement of the gravitational redshift of $\gamma$-ray 
lines emitted as a result of spallation processes on the surface of an 
acreting neutron star (Bildsten, Salpeter \& Wasserman 1993).

\noindent (ix) Time-resolved spectroscopy of X-ray bursts, in principle,
allows the derivation of constraints on the mass-radius relation of neutron 
stars (see Lewin, Van Paradijs \& Taam 1993, for a detailed discussion). 
The practical value of this method is limited by the systematic uncertainty 
in the interpretation of the burst spectra (Van Paradijs et al. 1991) and 
the possibility that the burst emission is not uniformly distributed 
across the neutron star surface (see L. Bildsten's contribution to this 
Volume). 

\subsection{Mass determinations for binary radio pulsars}

\subsubsection{Relativistic effects}

Radio pulse arrival times can be measured with exquisite accuracy (to
several tens of ns), and at this level of accuracy several relativistic effects 
become strongly detectable. These have been conveniently described by
Taylor \& Weisberg (1989), based on a theoretical formalism developed
by Damour \& Deruelle (1986), for the case that general relativity is
explicitly assumed to be valid. I refer the reader to these two papers
for details, and limit myself here to a brief listing of the effects,
as they have been applied so far to several binary radio pulsars.

By far the largest part of the pulse arrival time variations is the
light travel time across the orbit, which Taylor \& Weisberg describe
by the ``Roemer delay'' parameter $a_{\rm R}$ (close to the classical
semi-major axis of the orbit), which is related to the masses $m_{\rm p}$ 
and $m_{\rm c}$ of the pulsar and its companion,
respectively, their sum $M$, and the orbital period $P$, by:

\begin{equation}
{{a_{\rm R}^3}\over {P^2}} = {{GM}\over {4\pi^2}}\Bigl \lbrack 1 +
({{m_{\rm p}m_{\rm c}}\over {M^2}} - 9) {{GM}\over{2a_{\rm R}c^2}}
\Bigr \rbrack ^2
\end{equation}

The orbit precesses, at a rate measured by the quantity $k$, defined
by $\dot \omega = 2\pi k/P$, where $k$ is given by:

\begin{equation}
k = {{3GM}\over{c^2 a_{\rm R} (1-e^2)}}
\end{equation}

\noindent Here $\omega$ is the angle between the line of nodes and the 
direction toward periastron (as measured in the orbital plane), and $e$ is the
orbital eccentricity. 

The variation along the orbit of the
gravitational redshift and time dilation (Einstein delay) is given by
the quantity $\gamma$, related to the system parameters by: 

\begin{equation}
\gamma = {{ePGm_{\rm c}(m_{\rm p} + 2m_{\rm c})}\over {2\pi c^2 a_{\rm R}M}}
\end{equation}

Due to the emission of gravitational radiation the orbit decays; the
rate of decrease of the period is given by

\begin{equation}
\dot P = -{{192\pi}\over {5c^5}}\, ({2\pi G}/ {P})^{5/3}~
f(e)\,m_{\rm p}m_{\rm c}M^{-1/3}
\end{equation}

\noindent where

\begin{equation}
f(e) = \Bigl \lbrack 1 + {{73}\over {24}}e^2 + {{37}\over {96}}e^4
\Bigr \rbrack (1-e^2)^{-7/2}
\end{equation}

Finally, the pulses may show a measurable Shapiro delay, which
reflects that near the pulsar companion the path along which the
pulsar signal travels is curved. The shape of the orbital phase
dependence of this delay is characterized by two quantities $s$ and
$r$, given by

\begin{equation}
s = {\rm sin}\,i,~~~r = Gm_{\rm c}/c^3
\end{equation}

In the case of the Hulse-Taylor pulsar, PSR\,B1913+16, all above
effects have been measured, each of which provides a different
constraint on $m_{\rm p}$ and $m_{\rm c}$. As shown by
Taylor \& Weisberg (1989), this overdetermined set of constraints leads to
one consistent solution ($m_{\rm p} = 1.442 \pm 0.003$\,M$_{\odot}$, 
$m_{\rm c} = 1.386 \pm 0.003$\,M$_{\odot}$), showing that to the accuracy at
which the test can be performed, general relativity provides a
consistent description of this system.

Wolszczan (1991) measured the periastron advance and the Einstein
delay for the double neutron star system PSR\,B1534+12; adding the
constraint ${\rm sin}\, i \leq 1$ he derived $m_{\rm p} = 1.32 \pm
0.03$~M$_{\odot}$, and  $m_{\rm c} = 1.36 \pm 0.03$~M$_{\odot}$. 
Arzoumanian (1995) detected also orbital decay and Shapiro delay, and 
derived substantially improved masses for this system.

For PSR\,B1855+09, which is seen almost edge-on, Ryba \& Taylor (1991)
could measure the Shapiro delay, which directly gives the orbital
inclination and the companion mass ($m_{\rm c} =
0.233^{+0.026}_{-0.017}$\,M$_{\odot}$); together with the mass
function this gives $m_{\rm p} = 1.27^{+0.23}_{-0.15}$\,M$_{\odot}$. The
companion is a low-mass white dwarf; theoretical mass estimates for
its mass (based on evolutionary scenarios for the formation of systems
like PSR\,B1855+09) agree with the measured value (Joss et al. 1987;
Savonije 1987).

\begin{table} 
\caption{Neutron star masses from relativistic effects on binary pulsar 
timing} 
\begin{center}
\begin{tabular}{llccc}
\hline \\
Name & Method$^{a}$ & $m_{\rm p}$ & $m_{\rm c}$   & Ref. \\ 
     &        &(M$_{\odot}$)& (M$_{\odot}$) & \\ 
\hline \\
J1518--4904 & R,$\dot \omega$   & $1.54\pm 0.22$ & $1.09\pm 0.19$ & [1] \\
B1534+12    & R,$\dot \omega$,E,G,S & $1.338\pm 0.012$ & $1.341\pm 0.012$ & [2,7] \\
B1802--07   & R,$\dot \omega$  &$1.28\pm 0.16$&$0.35\pm 0.07$ & [3,7] \\
B1855+09    & R,S    &$1.27\,(+0.23,-0.15)$&$0.233\,(+0.026,-0.017)$& [4] \\
B1913+16    & R,$\dot \omega$,E,G,S &$1.442\pm 0.003$&$1.386\pm 0.003$ & [5] \\  
B2127+11C   & R,$\dot \omega$,E,G & $1.350\pm 0.040$ & $1.363\pm 0.060$& [6] \\
B2303+46    & R,$\dot \omega$   & $1.20\pm 0.26$ & $1.40\pm 0.24$ & [3,7] \\ \hline

\end{tabular}
\end{center}
\footnotesize
$^{a}$ R = Roemer delay; $\dot \omega$ = periastron advance; E = Einstein 
delay; G = orbital decay by gravitational radiation; S = Shapiro delay. 
References: 
[1] Nice et al. (1996);
[2] Wolszczan (1991);
[3] Thorsett et al. (1993);
[4] Ryba \& Taylor (1991);
[5] Taylor \& Weisberg (1989);
[6] Deich \& Kulkarni (1996);
[7] Arzoumanian (1995).

\normalsize

\end{table}

For PSR\,B2127+11C in the globular cluster M15 Deich \& Kulkarni (1996) 
measured the orbital precession, the Einstein delay and the orbital decay 
rate, from which they derived $m_{\rm p} =
1.350 \pm 0.040$\,M$_{\odot}$,  $m_{\rm c} = 1.363 \pm 0.040 $\,M$_{\odot}$.

Thorsett et al. (1993) measured the rate of periastron advance for
PSR\,B1802--07 and PSR\,B2303+46, from which they derive total system
masses $M$ of $1.7 \pm 0.4$\,M$_{\odot}$ and $2.53 \pm
0.08$\,M$_{\odot}$, respectively. Combining this with the observed
mass functions, and assuming a probable inclination range, they derive
$m_{\rm p} = 1.4^{+0.4}_{-0.3}$\,M$_{\odot}$,  $m_{\rm c}
=0.33^{+0.13}_{-0.10}$\,M$_{\odot}$, for PSR\,B1802--07, and  $m_{\rm p} =
1.16 \pm 0.28$\,M$_{\odot}$,  $m_{\rm c} = 1.37\pm 0.24 $\,M$_{\odot}$, for
PSR\,B2303+46. Improved masses for both systems have been derived by 
Arzoumanian (1995).

Nice et al. (1996) measured the rate of periastron advance and the mass 
function for PSR\,J1518+4904. Following the same argument as Thorsett 
et al. (1993) one finds for this system  $m_{\rm p} =
1.54 \pm 0.22$\,M$_{\odot}$,  $m_{\rm c} = 1.09\pm 0.19 $\,M$_{\odot}$.

\subsubsection{Non-relativistic mass determinations}

Van Kerkwijk et al. (1996) combined the pulse delay curves of the 
millisecond pulsar 
PSR\,J1012+5307  with the radial-velocity curve of its white-dwarf
companion.  This led to an accurately determined mass ratio $m_{\rm
p}/m_{\rm c} = 13.3 \pm 0.7$. For a given white-dwarf composition the 
surface gravity acceleration, as inferred  from model atmosphere fits
to the profiles of the Balmer absorption lines in the white-dwarf
spectrum, uniquely determine the white-dwarf mass, for which Van
Kerkwijk et al. derive a value of $0.16 \pm 0.02\,
(1\sigma)$\,M$_{\odot}$. Combining this with the mass ratio and the
mass function they find that the pulsar mass is in the range 1.5 to
3.2~M$_{\odot}$ (95\% confidence).

\subsection{Mass determinations for neutron stars and black holes in 
X-ray binaries}

\subsubsection{Mass function}

In determining the mass of an X-ray source, using Newtonian  
effects only, the fundamental quantity is the mass function $f_{\rm opt}(M)
$ which is determined from the orbital period, $P_{\rm orb}$, and the 
amplitude, $K_{\rm opt}$, of the radial-velocity variations of the mass 
donor by

\begin{equation}
f_{\rm opt}(M) \equiv M_{\rm X}^3~{\rm sin}^3~i / (M_{\rm X}+M_2)^2
= {{K^3_{\rm opt}~P_{\rm orb}} \over {2\pi G}}
\end{equation}

The corresponding quantity $f_{\rm X}(M)$ can be determined for binary 
X-ray pulsars:

\begin{equation}
f_{\rm X}(M) \equiv M_2^3~{\rm sin}^3~i / (M_{\rm X}+M_2)^2
= {{4 \pi^2~(a_{\rm X}{\rm sin}~i)^3} \over {GP^2_{\rm orb}}}
\end{equation}

\noindent (The connection to observational parameters is written
differently,  since in the  case of X-ray pulsars the observed
quantities are usually pulse arrival  times, whereas from optical
spectra one measures radial velocities.)

Incomplete but occasionally extremely useful information may be obtained 
from a measurement of only one mass
function, since it gives a lower limit to the mass of the companion of the 
star whose orbital motion is measured. As emphasized by McClintock \& 
Remillard (1986) this is of great importance in distinguishing black holes 
from neutron stars in X-ray binaries.

If both mass functions can be measured, their ratio immediately gives the 
mass ratio $q \equiv M_{\rm X}/M_2 = f_{\rm opt}(M)/f_{\rm X}(M)$, and both 
masses are then determined separately, up to a factor sin$^3~i$.

\begin{equation}
M_{\rm X} {\rm sin}^3~i = f_{\rm opt}(M)\,(1+q^{-1})^2
\end{equation}
\begin{equation}
M_{\rm opt} {\rm sin}^3~i = f_{\rm X}(M)\,(1+q)^2
\end{equation}

\subsubsection{Inclination angle}

To complete the mass determination one needs the orbital inclination
$i$, for whose determination several methods are available, at least in 
principle: (i) X-ray eclipse durations, (ii) optical light curves, 
and (iii) polarization variations.

For a spherical companion star with radius $R$ and a circular orbit
(separation $a$, period $P$) the duration of the eclipse of a
point-like X-ray source is related to $i$ by the expression:

\begin{equation}
\bigl (R/a \bigr )^2 = {\rm cos}^2\,i + {\rm sin}^2\,i~{\rm
sin}^2\,\theta_{\rm e}
\end{equation}

\noindent Here $\theta_{\rm e} = 2\pi t_{\rm e}/P$, with $t_{\rm e}$
half the duration of the eclipse. If the relative size of the
companion is known, $i$ is a function of $\theta_{\rm e}$ only. Direct
estimates of $R$ from the spectrum and luminosity class are not
accurate enough to be useful. In general, the companion will not be
spherical, due to the gravitational perturbation of the compact star.
The relative size of the primary can be expressed as a function of
the mass ratio $q$ and a dimensionless potential parameter $\Omega$,
which is a measure of the extent to which the companion fills its Roche lobe.
An observed eclipse duration then determines a relation between $q$,
$\Omega$, and $i$. The eclipse duration may be affected by absorbing
effects on the X-rays by a stellar wind from the companion (see, e.g., Woo 
et al. 1995). 

Electron scattering of originally unpolarized light in close-binary
stars may yield a net polarization which varies with orbital phase,
due to the deformation from spherical symmetry of the system (e.g.,
deformation of the companion star, presence of an accretion disk). 
Under rather general conditions the
fundamental and first harmonic (in orbital frequency) of the
variations of the Stokes parameters $Q$ and $U$ (see Tinbergen 1996)
describe ellipses in the $(Q,U)$ plane, whose eccentricity $e$ is
related to the orbital inclination by $e = {\rm sin}\,i$ for the
fundamental, and $e = {\rm sin}^2\,i/(1 + {\rm cos}^2\,i)$ for the
first harmonic (Brown et al. 1978; Rudy \& Kemp 1978 ; Milgrom 1979). 
Polarization variations may therefore provide a
measurement of $i$. The method has been applied to several HMXB, e.g., Cyg 
X-1 (Dolan \& Tapia 1989); for references to early work, see Van Paradijs 
(1983). 

As mentioned in Section 2.1.1, many HMXB with an evolved companion show
moderate (up to $\sim 10$\,\%)  optical brightness variations, with
two approximately equal maxima, and two different minima, which occur
at the quadratures and conjunctions, respectively. These so-called
``ellipsoidal'' light curves are caused by the rotational and tidal 
distortion of the companion star, and a non-uniform surface brightness
distribution (``gravity darkening''). The double-waved shape of the
light curve reflects the pear-like shape of the companion: near
conjunctions the projected stellar disk is smallest, near quadratures
largest.  For assumed co-rotation of the companion star, its
distortion is determined by the mass ratio $q$, and by the dimensionless
potential parameter $\Omega$ (see above); Roche lobe filling of the companion
corresponds to a ($q$ dependent) critical value of $\Omega$. The
distortion, and therefore the shape and amplitude of the ellipsoidal
light curve, are determined by $q$ and $\Omega$, and furthermore by 
$i$. Thus, in principle, the optical light curve can provide a
relation between $q$, $\Omega$ and $i$. Together with other
constraints on these parameters (eclipse duration, $q$ from two
observed mass functions) this may lead to a solution for the masses of
both components of the binary (see Van Paradijs 1983, for extensive 
references to early work on light curve modelling of HMXB).

The analysis by Tjemkes et al. (1986) showed that for well-studied HMXB
with evolved companions  the optical light curves may be reproduced if
the best-known system parameters are used. Conversely, however, 
X-ray heating of
the companion, the presence of an accretion disk, and intrinsic
variability of the companion  have a significant effect on the light
curve, to the extent that it is difficult to derive significant
constraints on these systems from an analysis of their light curves.
For Be/X-ray binaries, irregular brightness variations related to 
equatorial mass shedding in general dominate any orbital brightness 
variations.

Ellipsoidal light curves have been observed for several soft X-ray
transients in quiescence, whose optical light is then dominated by the
companion star. Since these companions  fill their Roche lobes, the
potential parameter $\Omega$ is determined by $q$, and therefore the
light curve depends on $q$ and $i$ only. If needed, the generally
small contribution from the quiescent accretion disk may be estimated
from a study of the broad-band energy distribution, and corrected for.
The contribution of the disk diminishes with increasing wavelength, and 
recent orbital light curve studies are therefore preferentially carried 
out in the infra-red (Casares et al. 1993; Beekman et al. 1996; Shahbaz et 
al. 1997).
The ellipsoidal light curves of SXT in quiescence have played an
important role in the mass estimates of black holes. However, one should
not ignore Haswell's (1995) emphasis that also for quiescent SXT
the ellipsoidal light curves may be affected by systematic effects,
possibly caused by variable contributions from an accretion disk,
analogius to the superhumps in the SU UMa subgroup of the cataclysmic
variables. 

\subsubsection{Additional Constraints}
            
In addition to the above standard ingredients of a mass determination
for X-ray binaries two additional pieces of information may be used.
The first applies when the source distance is known. Since the
apparent magnitude of the companion star (corrected for interstellar
extinction) and its spectral type together determine its angular
radius, for these sources one can estimate the radius, $R_2$, of the
companion star, and this constrains the relation between $M_{\rm X}$
and $M_2$. This was used by Gies \& Bolton (1986) in their 
analysis of Cyg X-1. In
case the companion star fills its Roche lobe, the constraint is
simple: for such stars the density is determined by the orbital period
alone (see, e.g., Frank et al. 1992), and $M_2$ follows 
immediately from $R_2$. In the case of
the SXT Cen X-4 (whose compact star is a neutron star) this served to
show that the companion mass is extremely low, less than
0.2\,M$_{\odot}$ (Chevalier et al. 1989; McClintock \& Remillard 1990). 

If the orbital angular momentum is parallel to that of the companion
star (and therefore of the matter flowing through the accretion disk), 
both $K_{\rm opt}$ and the observed rotational velocity $V_{\rm
rot}\,{\rm sin}\,i$ of the secondary have been decreased by the same
projection factor sin\,$i$. Therefore, their ratio is not affected. As
shown by Gies \& Bolton (1986) this constrains the relation between $M_{\rm X}$
and $M_2$. In case the companion star co-rotates at the same angular
velocity as the orbit, this constraint is an estimate of the mass
ratio. Using the expressions for the radius of the Roche lobe by
Paczynski (1971), one obtains (Kuiper et al. 1988; Wade \& Horne 1988):

\begin{equation}
\Bigl ( V_{\rm rot}\,{\rm sin}\,i/K_{\rm opt} \Bigr ) = 
0.462~q^{-1/3}\, (1+q)^{2/3}
\end{equation}

The cancellation of the sin\,$i$ factor has also been applied to the
rotational velocity at some radial distance in the accretion disk, as
inferred from emission line profiles.  Warner (1976)
applied this to cataclysmic variables, and Johnston et al. (1990) to the 
SXT A0620--00. Since relatively litle is known about the emission line 
structure of accretion disks, the 
interpretation of the $q$ values inferred from this method is somewhat 
uncertain. 

\begin{table} 
\caption{Neutron Star Masses: X-ray Binaries and Binary Radio 
Pulsars\hspace{2cm}}
\begin{center}
\begin{tabular}{lcccc}
\hline \\
Name & $i$ & $M_{\rm X}$ & $M_{\rm comp}$ & Ref. \\ 
     & ($^{\circ}$) & (M$_{\odot}$) & (M$_{\odot}$) & \\ \hline
{\it HMXB }\\
Vela X-1     & $>74$      & 1.88(+0.69,--0.47) & 23.5(+2.2,--1.5) & [1] \\
4U\,1538--52 & 68(+9,--8) & 1.06(+0.41,--0.34) & 16.4(+5.2,--4.0) & [1] \\
SMC X-1      & 70(+11,--7)& $1.6\pm 0.1$       & $17.2\pm 0.6$    & [2] \\
LMC X-4      & 65(+7,--6) & 1.47(+0.44,--0.39) & 15.8(+2.3,--2.0) & [1] \\
Cen X-3      & $>66$      & 1.09(+0.57,--0.52) & 18.4(+4.0,--1.8) & [1] \\
Her X-1$^a$  & $>79$      & 1.47(+0.23,--0.37) & 2.32(+0.16,--0.29) & [1] \\
Her X-1      & $>79$      & $1.5\pm 0.3$       & $2.3\pm 0.3$     & [3] \\ 
\hline
{\it LMXB } \\
Cen X-4      &  30-37     & 1.1-1.9            & $<0.2$           & [4,5] \\
4U\,1626--67$^a$ &  9-36  & 1.8(+2.8,--1.3)    & $<0.5$           & [6] \\
Cyg X-2      &  $<73$     & $>1.42(\pm 0.08)$  & $>0.47(\pm 0.03)$ & [7] \\ 
4U\,2129+47  &  $>70$     & $0.6\pm 0.2$       & $0.4\pm 0.2$     &  [8] \\
\hline
{\it LMBP} \\
J1012+5307   &            & 1.5-3.2            & $0.16\pm 0.02$   &  [9] \\ \hline

\end{tabular}
\end{center}
\footnotesize
$^a$ From Doppler shifted optical pulsations. \\
References: 
[1] Van Kerkwijk, Van Paradijs \& Zuiderwijk (1995);
[2] Reynolds et al. (1993);
[3] Reynolds et al. (1997); 
[4] Chevalier et al. (1989)
[5] McClintock \& Remillard (1990);
[6] Middleditch et al. (1981);
[7] Casares, Charles \& Kuulkers (1997); 
[8] Horne, Verbunt \& Schneider (1986); 
[9] Van Kerkwijk, Bergeron \& Kulkarni 1996). 
\end{table}

\normalsize

\subsubsection{Optical pulsations}

Optical pulsations have been detected from Her X-1 and 4U\,1626--67.
These pulsations arise from the reprocessing of pulsed X rays in the
companion star and the accretion disk, and can be used to study the
orbital parameters of these systems.

A detailed analysis of the optical pulsations of Her X-1 was made by
Middleditch \& Nelson (1976). The orbital motion of the neutron star
is known from the Doppler shifts of the X-ray pulse arrival times.
Optical pulsations in phase with the X-ray pulsations are present, but
also optical pulsations with a slightly different frequency. The
former arise from reprocessing in the accretion disk, the latter from
the surface of the companion; their frequency difference is just the 
orbital frequency (beat frequency relation). 
The pronounced variation with orbital
phase of the amplitude of the optical pulsations indicates that the
companion is  non-spherical and fills its Roche lobe. By assuming that
across the surface of the  Roche lobe filling companion the X-ray
reprocessing time is constant Middleditch \& Nelson used the
Doppler shift information of these optical pulsations to estimate the
mass ratio, which together with the eclipse duration and the X-ray
mass function leads to $M_{\rm X} = 1.30 \pm 0.14$\,M$_{\odot}$ and 
$M_{\rm opt} = 2.18 \pm 0.11$\,M$_{\odot}$. These values are consistent 
with those obtained from the radial-velocity curve of the late A-type 
companion star (Reynolds et al. 1997).

Doppler shifts of the 7.7 s X-ray pulsations of 4U\,1626--67 have not
been detected so far: $a_{\rm X}\,{\rm sin}\,i < 10$ ms (Levine et al.
1988). Optical pulsations, arising from reprocessing of X rays in the
disk and in the companion star were detected  by Middleditch et al.
(1981; see also Chakrabarthy et al. 1997). 
>From a detailed modelling of these pulsations, similar to that
for Her X-1, they derived  $M_{\rm X} =
1.8^{+2.9}_{-1.3}$\,M$_{\odot}$ and  $M_2 < 0.5$\,M$_{\odot}$.

\subsection{Summary of mass determinations of neutron stars and black 
holes}

The results of the mass estimates, described in Sections 4.2 and 4.3 are 
summarized in Table 2 (black holes in X-ray binaries), Table 4 
(radio pulsars), and Table 5 (neutron stars in X-ray 
binaries). From these tables one may draw the conclusion that with few 
exceptions neutron star masses are consistent with a relatively narrow 
mass range near the ``canonical'' value of 1.4\,M$_{\odot}$. The 
anomalously low value of the neutron star mass for 4U\,2129+47 (Horne, 
Verbunt \& Schneider 1986) may be due to contaminating light by a star that 
may be a distant outer companion in a triple or a foreground star (Garcia 
et al. 1989; private communication from F. Verbunt). Vela X-1 and 
Cyg X-2 present interesting examples of substantially higher neutron star 
masses, but the possible presence of systematic effects on these mass 
determinations has to be investigated before the consequence, i.e., that 
very soft equations of state are excluded, can be accepted.

\vspace*{1.0cm}
\noindent{\bf Acknowledgements}
I thank Dipankar Bhattacharya, Chryssa Kouveliotou, Walter Lewin, Ed van 
den Heuvel, Michiel van der Klis, Ralph Wij\-ers, and Nan Zhang for many 
discussions on topics covered in this review. I thank Frank van der Hooft, 
Marnix Witte and Erik Kuulkers for their help in preparing the manuscript.

\end{document}